\documentclass[aps,prd,preprint,tightenlines,groupedaddress,nofootinbib,showpacs,byrevtex]{revtex4}
\usepackage{amssymb,latexsym}
\usepackage{amsmath,amsbsy}
\usepackage{epsfig,bm}
\usepackage{graphicx,comment}
\DeclareMathOperator{\tr}{tr}

\begin{document}
\def\a{{\alpha}}
\def\b{{\beta}}
\def\d{{\delta}}
\def\D{{\Delta}}
\def\e{{\varepsilon}}
\def\g{{\gamma}}
\def\G{{\Gamma}}
\def\k{{\kappa}}
\def\l{{\lambda}}
\def\L{{\Lambda}}
\def\m{{\mu}}
\def\n{{\nu}}
\def\o{{\omega}}
\def\O{{\Omega}}
\def\S{{\Sigma}}
\def\s{{\sigma}}
\def\th{{\theta}}

\def\ol#1{{\overline{#1}}}

\def\Dslash{D\hskip-0.65em /}

\def\CPT{{$\chi$PT}}
\def\QCPT{{Q$\chi$PT}}
\def\PQCPT{{PQ$\chi$PT}}
\def\tr{\text{tr}}
\def\str{\text{str}}
\def\diag{\text{diag}}
\def\order{{\mathcal O}}

\def\cC{{\mathcal C}}
\def\cB{{\mathcal B}}
\def\cT{{\mathcal T}}
\def\cQ{{\mathcal Q}}
\def\cL{{\mathcal L}}
\def\cO{{\mathcal O}}
\def\cA{{\mathcal A}}
\def\cQ{{\mathcal Q}}
\def\cR{{\mathcal R}}
\def\cH{{\mathcal H}}
\def\cW{{\mathcal W}}
\def\cM{{\mathcal M}}
\def\cJ{{\mathcal J}}

\def\eqref#1{{(\ref{#1})}}

%\preprint{DUKE-TH-04-XXX}
 
\title{Baryon masses at $\cO(a^2)$ in chiral perturbation theory}
\author{ Brian C.~Tiburzi}
\email[]{bctiburz@phy.duke.edu}
\affiliation{Department of Physics\\
Duke University\\
P.O.~Box 90305\\
Durham, NC 27708-0305}

\date{\today}


\begin{abstract}
The chiral Lagrangian for the Symanzik action through $\cO(a^2)$ for baryons is obtained.
We consider two flavor unquenched and partially quenched lattice theories, allowing for mixed 
actions in the latter. As an application, we calculate masses to $\cO(a^2)$ for the 
nucleons and deltas, and investigate the corrections due to the violation of $O(4)$ rotational invariance. 
These results are contrasted with those in the meson sector for lattice 
simulations using mixed and unmixed actions of Wilson and Ginsparg-Wilson quarks.
\end{abstract}

\maketitle

\section{Introduction}
Lattice gauge theory provides first principles calculations of strong 
interaction physics, where QCD is non-perturbative, and quarks and gluons are confined 
in color-neutral hadronic states. 
These calculations, however, are severely limited by available computing power, 
necessitating the use of quark masses $m_q$ that are much larger than those in reality. 
To make physical predictions, one must extrapolate from the quark masses used on the lattice to those of nature.
A model independent tool for this extrapolation is to study QCD at hadronic scales 
using its low-energy effective theory, 
chiral perturbation theory (\CPT).
Because \CPT\ provides a systematic expansion involving $m_q/\L_{QCD}$, 
one can understand how QCD observables 
behave, in principle, as functions of the quark mass. 
To address the quenched and partially quenched approximations employed by
lattice calculations, \CPT\ has been extended to quenched chiral 
perturbation theory (\QCPT)~% 
\cite{Morel:1987xk,Sharpe:1992ft,Bernard:1992mk,Labrenz:1996jy,Sharpe:1996qp}
and partially quenched chiral perturbation theory (\PQCPT)~%
\cite{Bernard:1994sv,Sharpe:1997by,Golterman:1998st,Sharpe:2000bc,Sharpe:2001fh,Sharpe:2003vy}.

While lattice calculations are limited to unphysically large quark masses, they are restricted further by
two additional parameters: the size of the lattice $L$, that is not considerably larger than the system under investigation; 
and the lattice spacing $a$, that is not considerably smaller than the relevant hadronic distance scale.
To address the issue of finite lattice spacing,
\CPT\ has been extended (following the earlier work of~\cite{Sharpe:1998xm,Lee:1999zx})
in the meson sector
to $\cO(a)$ for the Wilson action~\cite{Rupak:2002sm}, and for mixed lattice actions~\cite{Bar:2002nr}.
Corrections at $\cO(a^2)$ in \CPT\ have been pursued~\cite{Bar:2003mh,Aoki:2003yv}.
There have also been parallel developments in addressing lattice spacing artifacts 
in staggered \CPT\ for mesons~\cite{Aubin:2003mg,Aubin:2003uc,Sharpe:2004is},
and heavy mesons~\cite{Aubin:2004xd}; and in twisted mass 
QCD~\cite{Munster:2003ba,Munster:2004dj,Scorzato:2004da,Sharpe:2004ps,Sharpe:2004ny}.
Corrections to baryon observables in \CPT\ and \PQCPT, too, have been recently investigated~\cite{Beane:2003xv,Arndt:2004we}. 
To consider finite lattice spacing corrections, one must formulate the underlying lattice theory and match
the new operators that appear onto those in the chiral effective theory. This can be done by utilizing a dual 
expansion in quark mass and lattice spacing. For an overview, see~\cite{Baer:2004xp}. 
Following~\cite{Bar:2003mh,Beane:2003xv}, we assume a hierarchy of energy scales
\begin{equation}
m_q \ll \L_{QCD} \ll \frac{1}{a}
.\end{equation}
We shall further choose a power counting scheme in which
the small dimensionless expansion parameters are\footnote{%
In practice the power counting scheme and subsequent ordering of the dual expansion in 
quark mass and lattice spacing should be organized based the on the actual sizes of $m_q$ and $a$. 
}
\begin{equation} \label{eqn:pc}
\e^2 \sim 
\begin{cases}
 m_q/\L_{QCD}, \\
 a \, \L_{QCD}
\end{cases}
.\end{equation}
Thus we have a systematic way to calculate $a$-dependent corrections
in \CPT\ for the observables of interest. Such expressions allow one to perform the quark mass extrapolation
before the continuum extrapolation.

In this work we address the extension of \CPT\ and \PQCPT\ at finite lattice spacing to $\cO(a^2)$ in 
the baryon sector. Specifically we detail the operators which must be included to determine
the baryon masses to $\cO(\e^4)$ in the above power counting. Unlike the extension of \CPT\ and \PQCPT\ 
to $\cO(a^2)$ in the meson sector, the baryon sector suffers from a proliferation of new operators. 
Despite this fact, the number of free independent parameters entering expressions for the masses
is still relatively small. Moreover in contrast to the meson sector at $\cO(a^2)$, heavy baryon 
operators that break the $O(4)$ rotational symmetry of Euclidean space are required at this order in 
the chiral expansion. Such operators are required for particles that are heavy compared to $\L_{QCD}$.

This paper has the following organization. First in Sec.~\ref{s:sym}, we review the Symanzik Lagrangian
at $\cO(a^2)$ for a partially quenched, mixed lattice action, where the valence and sea quarks
are either Wilson or Ginsparg-Wilson fermions. We focus on the symmetries of the Symanzik Lagrangian because
these are essential in constructing the effective theory. Next in Sec.~\ref{s:mesons}, we review the finite lattice
spacing corrections at $\cO(a)$ in the meson sector of \PQCPT.  Higher-order corrections are not needed for 
baryon observables to the order we work. In Sec.~\ref{s:baryons}, we extend heavy baryon \PQCPT\ to $\cO(\e^4)$. 
This includes the addition of operators at $\cO(m_Q \, a)$ and $\cO(a^2)$. Applications of this development
are pursued in Sec.~\ref{s:mass}, where we obtain the lattice spacing corrections to the masses of nucleons and deltas. 
The unquenched two-flavor theory is addressed in Appendix~\ref{s:2}.
Corrections from $O(4)$ breaking operators are treated in detail for particles of spin less than two in Appendix~\ref{s:RPI}.
A summary (Sec.~\ref{s:summy}) highlights the lattice spacing corrections in the baryon sector. 
Here we contrast the number of independent parameters entering expressions at $\cO(\e^4)$ for baryon masses in the various
mixed and unmixed partially quenched theories, as well as in the unquenched theory.

\section{\label{sec:PQCPT}\PQCPT\ at $\cO(a^2)$}

To extend the baryon chiral Lagrangian to $\cO(a^2)$, we first review 
the Symanzik Lagrangian at $\cO(a^2)$ and the construction of \PQCPT\ in the meson sector. 
In this and the following sections, we consider a partially quenched theory with a mixed action. 
The result for unquenched simulations is contained in Appendix~\ref{s:2}.

\subsection{Symanzik Lagrangian}  \label{s:sym}

The Symanzik action is the continuum effective theory of the lattice 
action~\cite{Symanzik:1983dc,Symanzik:1983gh}. As such it is constructed from continuum 
operators based on the symmetries of the underlying lattice theory. The Symanzik 
Lagrangian is organized in powers of the lattice spacing $a$, namely
\begin{equation}
\cL = 
\cL^{(4)} 
+ 
a \, \cL^{(5)} 
+ 
a^2 \, \cL^{(6)} 
+ 
\dots \label{eq:syman}
,\end{equation}
where $\cL^{(n)}$ represents contributions for dimension-$n$ operators.\footnote{%
Not all of the $a$-dependence is parametrized in Eq.~\eqref{eq:syman}. The coefficients of 
terms in the Lagrangian $\cL^{(n)}$ depend upon the gauge coupling and thus can have a 
weak logarithmic dependence on $a$. Such dependence is beyond the scope of this work. 
} 
In the continuum limit, $a \to 0$, only the dimension-four operators survive.
The Symanzik Lagrangian for the Wilson action was discussed to $\cO(a^2)$ 
in~\cite{Sheikholeslami:1985ij} and the analysis to $\cO(a)$ was refined by~\cite{Luscher:1996sc}. 
Here we consider the general case of a mixed action in partially quenched QCD (PQQCD). Such an action
allows the valence and sea quarks to have different masses. Additionally the valence and sea quarks 
can be different types of lattice fermions. 
The mixed lattice action has the usual parity invariance, charge conjugation invariance and $SU(N_c)$ gauge symmetry. 
Accordingly the Symanzik Lagrangian $\cL$ respects these symmetries order-by-order in $a$. 
Because spacetime has been discretized, the $O(4)$ rotational symmetry of continuum Euclidean
field theory has been reduced to the hypercubic group.

The flavor symmetry group of the mixed lattice 
action $G$ is generally a direct product of flavor symmetry groups for the particular species of 
fermion at hand~\cite{Bar:2002nr}. The mixed action contains different forms of the Dirac 
operator for each species of fermion, thus there is no symmetry transformation between the 
valence and sea sectors.  At zero quark mass, the flavor symmetry group of the mixed lattice action is 
\begin{equation} 
G = G_{\text{valence}} \otimes G_{\text{sea}}
\label{eq:mixsym}
,\end{equation}
where for two quark flavors
\begin{equation}
G_{\text{valence}} = SU(2|2)_L \otimes SU(2|2)_R
,\end{equation}
and
\begin{equation}
G_{\text{sea}} = SU(2)_L \otimes SU(2)_R
.\end{equation}
For simplicity we have included the ghost quarks in the valence sector. 
The quark mass term breaks these respective chiral symmetries in the valence and sea sectors 
down to vector symmetries.

For Wilson fermions~\cite{Wilson:1974sk}, the breaking of chiral symmetry persists even at zero quark mass 
due to lattice discretization effects. These effects can be tamed to enter at $\cO(a^2)$ by the so-called $\cO(a)$ improvement.  
For fermions satisfying the Ginsparg-Wilson relation~\cite{Ginsparg:1982bj}
(e.g., Kaplan fermions~\cite{Kaplan:1992bt}, or overlap fermions~\cite{Narayanan:1993ss}), 
the chiral symmetry on the lattice at zero quark mass remains exact at finite $a$~\cite{Luscher:1998pq}.
For a recent review of chiral symmetry in lattice theories, see~\cite{Chandrasekharan:2004cn}. 
We have left unspecified which species lives in which sector of the theory. As a result of this generality, 
we are building the Symanzik Lagrangian for the four possible combinations of valence and sea fermions.
Part of this generality is an academic pursuit because we do not anticipate lattice calculations employing 
Wilson valence quarks in a Ginsparg-Wilson sea, whereas GW valence quarks with dynamical Wilson fermions
is a scenario with potential computational benefits. It is likely, however,  that  
Wilson quark masses may never reach the chiral regime~\cite{Beane:2004ks} in which case one must deal with alternate methods
to modify~\cite{Leinweber:1999ig} or to improve~\cite{Bernard:2003rp} baryon \CPT.
These reservations aside, our goal is to contrast the situation at $\cO(a^2)$
in the baryon sector with that of the mesons. We will find comparatively that the mesons are rather
special with respect to lattice spacing corrections in \CPT.

Having discussed the symmetries of $\cL$, we now turn to the contributions near the continuum limit.
The terms of the $\cL^{(4)}$ Lagrangian are the most familiar and so we discuss them first. 
At dimension four, we have the familiar kinetic and Dirac mass terms for the quarks. To be specific, 
we work in a partially quenched theory for two light flavors. 
The quark part of the leading-order Symanzik Lagrangian reads
\begin{equation}\label{eqn:LPQQCD}
\cL^{(4)}
=
\ol Q \, \Dslash \, Q  +  \ol Q \, m_Q \, Q
.\end{equation}
The six quarks of PQQCD are in the fundamental representation of
the graded group $SU(4|2)$%
~\cite{BahaBalantekin:1981kt,BahaBalantekin:1981qy,BahaBalantekin:1982bk}
and appear in the vector
\begin{equation}
  Q=(u,d,j,l,\tilde{u},\tilde{d})^{\text{T}}
,\end{equation}
which obeys the graded equal-time commutation relation
\begin{equation} \label{eqn:commutation}
  Q^\a_i({\bf x}){Q^\b_j}^\dagger({\bf y})
  -(-1)^{\eta_i \eta_j}{Q^\b_j}^\dagger({\bf y})Q^\a_i({\bf x})
  =
  \d^{\a\b}\d_{ij}\d^3({\bf x}-{\bf y})
,\end{equation}
where $\a$ and $\b$ are spin, and $i$ and $j$ are flavor indices.
The vanishing graded equal-time commutation relations can be written analogously.
The grading factor 
\begin{equation}
   \eta_k
   = \left\{ 
       \begin{array}{cl}
         1 & \text{for } k=1,2,3,4 \\
         0 & \text{for } k=5,6
       \end{array}
     \right.
,\end{equation}
incorporates the different statistics of the quarks. 
The quark mass matrix in the isospin limit ($m_d = m_u$) of $SU(4|2)$ is given by 
\begin{equation}
  m_Q=\text{diag}(m_u, m_u, m_j, m_j, m_u, m_u)
.\end{equation}
In the limit $a \to 0$, and when $m_j=m_u$, one recovers the isospin limit of QCD.
The symmetry of $\cL^{(4)}$ in the zero mass limit is $SU(4|2)_L \otimes SU(4|2)_R$,   
i.e.~the effects of the mixed action do not show up in the effective theory at leading order. 
The mass term in $\cL^{(4)}$ explicitly breaks the graded chiral symmetry down to the graded vector symmetry 
$SU(4|2)_V$.

Let us next consider the dimension-five Lagrangian. After field redefinitions, it consists of only the Pauli term given by
\begin{equation} \label{eqn:Pauli}
\cL^{(5)} 
= 
c_{SW} \, \ol Q \, \sigma_{\mu \nu} G_{\mu \nu} w_Q \, Q
.\end{equation}
This term breaks chiral symmetry in precisely the same 
way as the quark mass term. 
The Sheikholeslami-Wohlert (SW)~\cite{Sheikholeslami:1985ij} 
coefficient is $c_{SW}$ and it is accompanied by the Wilson matrix $w_Q$
defined by
\begin{equation} \label{eqn:sw}
w_Q = \text{diag}
(w_v,w_v,w_s,w_s,w_v,w_v)
.\end{equation}
This matrix accounts for 
the chiral symmetry properties of the mixed action.  
If the quark $Q_i$ is a Wilson fermion, 
then $(w_Q)_i = 1$.  Alternately, if $Q_i$ is of the 
Ginsparg-Wilson (GW) variety then $(w_Q)_i = 0$. 
Since one expects simulations to be performed with 
valence quarks that are all of the same species as well as sea quarks 
all of the same species, we have labeled the entries in Eq.~\eqref{eqn:sw}
by valence ($v$) and sea ($s$) instead of flavor.\footnote{%
It is conceivable that three flavor simulations might employ Wilson up and down quarks but with a GW strange
quark in order to eliminate the rather large $\cO( a\, m_s)$ corrections. The development in this
work for partially quenched theories is general enough to handle this situation by a 
modification of Eq.~\eqref{eqn:sw}. I thank W.~Detmold for bringing this to my attention.}

To work at $\cO(a^2)$, we must consider the dimension-six Lagrangian. The exact form of $\cL^{(6)}$ 
is irrelevant in constructing the chiral effective theory. We need only know which symmetries 
are broken and how. We explicitly list terms of the Lagrangian when it is illustrative to do so. 
In describing these operators, we introduce the flavor matrix 
$\ol w_Q \equiv 1 - w_Q$, which projects onto the GW sector of the theory.

The terms of $\cL^{(6)}$ fall into five classes.
The first class of operators consists of  higher-dimensional quark bilinears. 
These operators do not break chiral symmetry but have the flavor symmetry 
of the mixed action, i.e.~there are distinct operators involving only the valence sector and only the sea sector.
Typical examples are $\ol Q \, \Dslash \, {}^3 w_Q \, Q$, and $\ol Q \, \Dslash \, {}^3 \ol w_Q \, Q$. 
In class two, there are quark bilinear operators that break chiral symmetry. Simple dimensional counting
indicates that there must be an $m_Q$ insertion for these operators to be in $\cL^{(6)}$. Examples include
$\ol Q \, m_Q \, D^2 w_Q \, Q$, and $\ol Q \, m_Q \, D^2 \ol w_Q \, Q$. 
Class three consists of all four-quark operators that do not break the chiral symmetry
of the valence and sea sectors. These class three operators
come in three forms due to the different sectors of the theory, e.g.~$(\ol Q \, w_Q \, \gamma_\mu Q)^2$, 
$(\ol Q \, \ol w_Q \, \gamma_\mu Q)^2$,  and $ (\ol Q \, w_Q \, \gamma_\mu Q) (\ol Q \, \ol w_Q \, \gamma_\mu Q)$. 
Class four operators are four-quark operators that break chiral symmetry. Since there are four fields, chiral 
symmetry is broken in a different way than the quark mass term. A typical class four operator is $(\ol Q \, w_Q \, Q)^2$. 
Unlike class three, there are no class four operators involving the flavor matrix $\ol w_Q$. 
Finally class five operators are those that break $O(4)$ rotation symmetry. There are two such operators 
for the mixed lattice action, namely $\ol Q \, w_Q \, \gamma_\mu D_\mu D_\mu D_\mu Q$, and 
$\ol Q \, \ol w_Q \, \gamma_\mu D_\mu D_\mu D_\mu Q$.  These classes are summarized in Table~\ref{t:class}.

\begin{table}
\caption{Description of the classes of operators in $\cL^{(6)}$. Each representative 
class of operator is classified as to whether it is bilinear, chiral symmetry breaking ($\chi$SB),
or $O(4)$ breaking. Operators in $\cL^{(6)}$ that are not bilinears are thus four-quark operators. 
Additionally the flavor matrices involved for the mixed action are listed.  
}
%\begin{ruledtabular}
\begin{tabular}{c | c c c c c c   }
Class & $\qquad$ Bilinear? $\quad$ & $\qquad \chi$SB? $\qquad$ & $O(4)$ breaking? & Flavor Structure  \\
\hline
$1$ & Yes & No  & No  & $w_Q$, $\ol w_Q$ & \\ 
$2$ & Yes & Yes & No  & $m_Q \otimes w_Q$, $m_Q \otimes \ol w_Q$ & \\ 
$3$ & No  & No  & No  & $w_Q \otimes w_Q$, $w_Q \otimes \ol w_Q$, $\ol w_Q \otimes \ol w_Q$ & \\ 
$4$ & No  & Yes & No  & $w_Q \otimes w_Q$ & \\ 
$5$ & Yes & No  & Yes & $w_Q$, $\ol w_Q$ & 
\end{tabular}
%\end{ruledtabular}
\label{t:class}
\end{table}

\subsection{Mesons} \label{s:mesons}

For massless quarks at zero lattice spacing,
the Lagrangian in Eq.~(\ref{eqn:LPQQCD}) exhibits a graded symmetry
$SU(4|2)_L \otimes SU(4|2)_R \otimes U(1)_V$ that is assumed 
to be spontaneously broken to $SU(4|2)_V \otimes U(1)_V$. 
The low-energy effective theory of PQQCD that results from 
perturbing about the physical vacuum is \PQCPT.
The pseudo-Goldstone mesons can be described at $\cO(\e^2)$ by a Lagrangian 
that accounts for the two sources of explicit chiral symmetry breaking:
the quark mass term in Eq.~\eqref{eqn:LPQQCD}, and the Pauli term in Eq.~\eqref{eqn:Pauli}
\cite{Sharpe:1998xm,Rupak:2002sm,Bar:2002nr}:
\begin{equation}\label{eqn:Lchi}
  {\cal L} =
   \frac{f^2}{8}
    \str\left(\partial_\mu\Sigma^\dagger \partial_\mu\Sigma\right)
    - \l_m\,\str\left(m_Q\Sigma^\dagger+m_Q^\dagger\Sigma\right)
    - a \l_a\,\str\left(w_Q\Sigma^\dagger+w_Q^\dagger\Sigma\right) 
\end{equation}
where
\begin{equation} \label{eqn:Sigma}
  \Sigma=\exp\left(\frac{2i\Phi}{f}\right)
  = \xi^2
,\end{equation}
\begin{equation}
  \Phi=
    \left(
      \begin{array}{cc}
        M & \chi^{\dagger} \\ 
        \chi & \tilde{M}
      \end{array}
    \right)
,\end{equation}
$f=132$~MeV, and the str() denotes a graded flavor trace.  
The $M$, $\tilde{M}$, and $\chi$ are matrices
of pseudo-Goldstone bosons 
and pseudo-Goldstone fermions,
see, for example,~\cite{Chen:2001yi}.
Expanding the Lagrangian in \eqref{eqn:Lchi} one finds that
to lowest order mesons with quark content $Q\bar{Q'}$
have mass\footnote{%
The quark masses $m_Q$ above are not those customarily used on the lattice~\cite{Sharpe:1998xm,Aoki:2003yv}, 
because one usually defines the quark mass in terms of a critical parameter for which the meson masses vanish.  
If this is indeed the way one defines the renormalized quark mass, 
then the parameter $\l_a$ in the Lagrangian can be set to zero and the issue of $a$-dependent 
loop-meson masses in baryonic observables will never confront us.  Allowing for other definitions 
of the lattice renormalized quark mass to avoid this fine-tuning, we keep $\l_a \neq 0$ and deal with the possibility that 
the loop-meson masses remain $a$-dependent. Notice this issue only arises for Wilson quarks.
}
\begin{equation}\label{eqn:mqq}
  m_{QQ'}^2=\frac{4}{f^2} \left[ \l_m (m_Q+m_{Q'}) + a \l_a (w_Q + w_{Q'}) \right]
.\end{equation}
The flavor singlet field is rendered heavy by the $U(1)_A$ anomaly
and has been integrated out in \PQCPT, however,
the propagator of the flavor-neutral field deviates from a simple pole 
form~\cite{Sharpe:2001fh}. 
For $a,b = u,d,j,l,\tilde u,\tilde d$, the leading-order $\eta_a \eta_b$ propagator is given by
\begin{equation}
{\cal G}_{\eta_a \eta_b} =
        \frac{i \epsilon_a \delta_{ab}}{q^2 - m^2_{\eta_a} +i\epsilon}
        - \frac{i}{2} \frac{\epsilon_a \epsilon_b \left(q^2 - m^2_{jj}
            \right) }
            {\left(q^2 - m^2_{\eta_a} +i\epsilon \right)
             \left(q^2 - m^2_{\eta_b} +i\epsilon \right)}\, ,
\end{equation}
where
\begin{equation}
\epsilon_a = (-1)^{1+\eta_a}
.\end{equation}
The flavor neutral propagator can be conveniently rewritten as
\begin{equation}
{\cal G}_{\eta_a \eta_b} =
         \e_a \d_{ab} P_a +
         \e_a \e_b {\cal H}_{ab}\left(P_a,P_b\right),
\end{equation}
where
\begin{eqnarray}
     P_a &=& \frac{i}{q^2 - m^2_{\eta_a} +i\e},\ 
     P_b = \frac{i}{q^2 - m^2_{\eta_b} +i\e},\, 
\nonumber\\
\nonumber\\
     {\cal H}_{ab}\left(A,B\right) &=& 
           -\frac{1}{2}\left[
             \frac{m^2_{\eta_a} - m^2_{jj}}{m^2_{\eta_a} - m^2_{\eta_b}}
                 A
            -\frac{m^2_{\eta_b} - m^2_{jj}}{m^2_{\eta_b} - m^2_{\eta_a}}
                 B \ \right].
\label{eq:Hfunction2}
\end{eqnarray}

At $\cO(\e^4)$, one has contributions to the meson Lagrangian from
operators of $\cO(p^4)$, $\cO(p^2 \, m_Q)$, and $\cO(m_Q^2)$. These are
the Gasser-Leutwyler terms. Additionally there are terms of $\cO( p^2 \, a)$, 
$\cO(m_Q \, a)$, and $\cO(a^2)$ that are generalizations of the Gasser-Leutwyler
terms for the Symanzik lattice action.  These have been determined 
in~\cite{Bar:2003mh,Aoki:2003yv} and we do not duplicate them here. Since our concern 
does not lie in the meson sector, mesons will only enter via loop calculations. Retaining
the meson masses to $\cO(\e^4)$ in a typical baryon calculation leads to corrections of 
$\cO(\e^5)$ or higher. These are beyond the order we work, thus the 
Lagrangian in Eq.~\eqref{eqn:Lchi} is sufficient for our purposes.

\subsection{Baryons} \label{s:baryons}

Having reviewed the Symanzik Lagrangian through $\cO(a^2)$ and the relevant pieces of meson \PQCPT\ at finite $a$, 
we now extend \PQCPT\ in the baryon sector to $\cO(a^2)$. First let us detail the situation at $\cO(a)$. 
In $SU(4|2)$ \PQCPT, the spin-$\frac{1}{2}$ baryons 
are embedded in the $\bf{70}$-dimensional super-multiplet $\cB^{ijk}$, that
contains the nucleons, while the spin-$\frac{3}{2}$ baryons
are embedded in the $\bf{44}$-dimensional super-multiplet $\cT_\mu^{ijk}$, that 
contains the deltas~\cite{Labrenz:1996jy,Beane:2002vq}. 
To $\cO(\e^2)$, the free Lagrangian for the $\cB^{ijk}$ and $\cT^{ijk}_\mu$ fields is given 
by~\cite{Labrenz:1996jy,Beane:2002vq,Beane:2003xv}
\begin{eqnarray} \label{eqn:L}
  {\mathcal L}
  &=&
   i\left(\ol\cB v\cdot{\mathcal D}\cB\right)
  -2\a_M\left(\ol\cB \cB{\mathcal M}_+\right)
  -2\b_M\left(\ol\cB {\mathcal M}_+\cB\right)
  -2\sigma_M\left(\ol\cB\cB\right)\str\left({\mathcal M}_+\right)
                              \nonumber \\
  &&
  - 2 \a_W\left(\ol\cB \cB{\mathcal W}_+\right)
  - 2 \b_W\left(\ol\cB {\mathcal W}_+\cB\right)
  - 2 \sigma_W\left(\ol\cB\cB\right)\str\left({\mathcal W}_+\right)
                              \nonumber \\
  &&+i\left(\ol\cT_\mu v\cdot{\mathcal D}\cT_\mu\right)
    +\D\left(\ol\cT_\mu\cT_\mu\right)
    +2\g_M\left(\ol\cT_\mu {\mathcal M}_+\cT_\mu\right)
    -2\ol\sigma_M\left(\ol\cT_\mu\cT_\mu\right)\str\left({\mathcal M}_+\right) 
				\nonumber \\
  &&
  + 2 \g_W\left(\ol\cT_\mu {\mathcal W}_+\cT_\mu\right)
  - 2 \ol\sigma_W\left(\ol\cT_\mu\cT_\mu\right)\str\left({\mathcal W}_+\right)
,\end{eqnarray}
where the mass operator is defined by
\begin{equation}
{\mathcal M}_\pm = \frac{1}{2}\left(\xi^\dagger m_Q \xi^\dagger\pm\xi m_Q \xi\right)
,\end{equation}
and the Wilson operator is defined by\footnote{%
Technically the SW coefficient $c_{SW}$, with its possible weak logarithmic dependence on $a$, 
also enters into the spurion construction. We are ignoring this dependence and do not distinguish between 
spurions of the same form which come from different operators in the Symanzik action largely because at $\cO(a^2)$ 
the analysis becomes extremely cumbersome.
}
\begin{equation} \label{eq:Wilson}
\mathcal{W}_\pm = \frac{a \L_{QCD}^2}{2} \left(\xi^\dagger w_Q \xi^\dagger\pm\xi w_Q \xi\right)
.\end{equation}
Here $\D \sim \e$ is the mass splitting between the $\bf{70}$ and $\bf{44}$ in the chiral limit.
The parenthesis notation used in Eq.~\eqref{eqn:L} is that of~\cite{Labrenz:1996jy} and is defined 
so that the contractions of flavor indices maintain proper transformations under chiral rotations.
Notice that the presence of the chiral symmetry breaking SW operator in Eq.~\eqref{eqn:LPQQCD} 
has lead to new $\cO(a)$ operators and new dimensionless constants $\a_W$, $\b_W$, $\sigma_W$, $\g_W$, 
and $\ol\sigma_W$ in Eq.~\eqref{eqn:L}. 
%Notice in our power counting $a \L_\chi^2 \sim m_Q$. 
The Lagrangian describing the interactions of the $\cB^{ijk}$ 
and $\cT_\mu^{ijk}$ with the pseudo-Goldstone mesons is
\begin{equation} \label{eqn:Linteract}
  {\cal L} =   
	  2 \a \left(\ol \cB S_\mu \cB A_\mu \right)
	+ 2 \b \left(\ol \cB S_\mu A_\mu \cB \right)
	- 2{\mathcal H}\left(\ol{\cT}_\nu S_\mu A_\mu \cT_\nu\right) 
    	+ \sqrt{\frac{3}{2}}\cC
  		\left[
    			\left(\ol{\cT}_\nu A_\nu \cB\right)+ \left(\ol \cB A_\nu \cT_\nu\right)
  		\right]  
.\end{equation}
The axial-vector and vector meson fields $A_\mu$ and $V_\mu$
are defined by: $ A_\mu=\frac{i}{2}
\left(\xi\partial_\mu\xi^\dagger-\xi^\dagger\partial_\mu\xi\right)$  
and $V_\mu=\frac{1}{2} \left(\xi\partial_\mu\xi^\dagger+\xi^\dagger\partial_\mu\xi\right)$.
The latter appears in  Eq.~\eqref{eqn:L} for the
covariant derivatives of $\cB_{ijk}$ and $\cT_{ijk}$ 
that both have the form
\begin{equation}
  ({\mathcal D}_\mu \cB)_{ijk}
  =
  \partial_\mu \cB_{ijk}
  +(V_\mu)_{il}\cB_{ljk}
  +(-)^{\eta_i(\eta_j+\eta_m)}(V_\mu)_{jm}\cB_{imk}
  +(-)^{(\eta_i+\eta_j)(\eta_k+\eta_n)}(V_\mu)_{kn}\cB_{ijn}
.\end{equation}
The vector $S_\mu$ is the covariant spin operator~\cite{Jenkins:1991jv,Jenkins:1991es,Jenkins:1991ne}.
The interaction Lagrangian in Eq.~\eqref{eqn:Linteract} also receives finite $a$ corrections. 
In calculating the octet and decuplet masses, however, these lead to effects that are of $\cO(\e^5)$ or higher.

At $\cO(\e^4)$, there are contributions to the \PQCPT\ Lagrangian from two insertions of the mass operator $\cM_+$,
and contributions from two insertions of the axial current $A_\mu$. 
The former contribute to the baryon masses at tree level, the latter at one-loop level. 
These operators have been written down in \cite{Walker-Loud:2004hf,Tiburzi:2004rh,Tiburzi:2005na}.
There are also operators with an insertion of $v \cdot A \otimes \cM_-$. These do not contribute to the masses at $\cO(\e^4)$. 
To extend baryon \PQCPT\ for mixed lattice actions to $\cO(\e^4)$, we must first include all higher-order operators that are 
linear in $a$. Secondly we must map the operators in $\cL^{(6)}$ onto the baryon sector. To achieve the former, 
we realize that the relevant operators can be formed from the existing $\cO(a)$ operators by insertion of the mass operator $\cM_+$ 
(insertion of a derivative is ruled out because the only possibility is $v \cdot D$ which can be eliminated using
the equations of motion~\cite{Arzt:1993gz}). For spin-$\frac{1}{2}$ baryons, there are eleven such operators
\begin{eqnarray}
\cL &=&
-\frac{1}{\L_\chi} 
\Bigg[ 
 b_1^{WM} (-)^{(\eta_i + \eta_j)(\eta_k + \eta_{k'})} \ol \cB {}^{kji} \{\cM_+, \cW_+ \}^{kk'} \cB^{ijk'} 
+ 
b_2^{WM} \ol \cB {}^{kji} \{\cM_+, \cW_+ \}^{ii'} \cB^{i'jk}  
\notag \\
&& + 
b_3^{WM} (-)^{\eta_{i'}(\eta_j + \eta_{j'})} \ol \cB {}^{kji} \cM_+^{ii'} \cW_+^{jj'} \cB^{i'j'k} 
+ 
b_4^{WM} (-)^{\eta_{i}(\eta_j + \eta_{j'})} \ol \cB {}^{kji} \cM_+^{jj'} \cW_+^{ii'} \cB^{i'j'k} 
\notag \\
&& + 
b_5^{WM} (-)^{\eta_j \eta_{j'} + 1} \ol \cB {}^{kji} 
\left(\cM_+^{ij'} \cW_+^{ji'} + \cW_+^{ij'} \cM_+^{ji'} \right) \cB^{i'j'k} 
\notag \\
&& +
b_6^{WM} \left( \ol \cB \cB \cM_+\right) \str (\cW_+) 
+ 
b_7^{WM} \left( \ol \cB  \cM_+ \cB \right) \str (\cW_+)
+ 
b_8^{WM} \left( \ol \cB \cB \right) \str ( \cW_+ \cM_+ ) 
\notag \\
&& + 
b_9^{WM} \left( \ol \cB \cB \right) \str( \cW_+ ) \str (\cM_+) 
+ 
b_{10}^{WM} \left( \ol \cB \cB \cW_+ \right) \str (\cM_+)
+ 
b_{11}^{WM} \left( \ol \cB \cW_+ \cB \right) \str (\cM_+)
\Bigg],
\notag \\
\label{eq:BAMs}
\end{eqnarray}
involving $\cW_+ \otimes \cM_+$. 
Additionally there are eleven analogous operators involving the operator combination $\cW_- \otimes \cM_-$, however
these terms do not contribute to the baryon masses at this order. 
Similarly the eight operators containing $v \cdot A \otimes \cW_-$
do not contribute to baryon masses to $\cO(\e^4)$.

For the spin-$\frac{3}{2}$ baryons, there are six operators at $\cO( m_Q \, a )$ that contribute to the baryon masses. 
These terms are  
\begin{eqnarray}
\cL &=&
 \frac{1}{\L_\chi} \Bigg[ 
t_1^{WM} \ol \cT {}^{kji}_\mu \{\cM_+, \cW_+ \}^{ii'} \cT^{i'jk}_\mu  
+
t_2^{WM} (-)^{\eta_{i'}(\eta_j + \eta_{j'})} \ol \cT  {}^{kji}_\mu \cM_+^{ii'} \cW_+^{jj'} \cT^{i'j'k}_\mu
\notag \\ 
&& + 
t_3^{WM} \left( \ol \cT_\mu  \cM_+ \cT_\mu \right) \str (\cW_+)
+ 
t_4^{WM} \left( \ol \cT_\mu \cT_\mu \right) \str ( \cW_+ \cM_+ ) 
\notag \\
&& + 
t_5^{WM} \left( \ol \cT_\mu \cT_\mu \right) \str( \cW_+ ) \str (\cM_+) 
+ 
t_{6}^{WM} \left( \ol \cT_\mu \cW_+ \cT_\mu \right) \str (\cM_+)
\Bigg].
\label{eq:TAMs}
\end{eqnarray}
Additionally there are six analogous terms involving the operator combination $\cW_- \otimes \cM_-$ and four terms containing
$v \cdot A \otimes \cW_-$,  but these operators do not 
contribute to the baryon masses at this order.

Next we must assess the contribution from the different classes of operators in $\cL^{(6)}$. 
Class one operators are quark bilinears that do not break chiral symmetry in the valence and sea sectors. 
These operators, however, break the full graded chiral symmetry down to the chiral symmetry group $G$
of the mixed lattice action. To describe such operators in the effective theory, 
we rewrite the class one operators of the Symanzik Lagrangian in a different form. Instead of describing the
bilinears in terms of $w_Q$ and $\ol w_Q$ matrices, we use the linear combinations $w_Q + \ol w_Q$ and 
$w_Q - \ol w_Q$. The former combination is just the identity and terms in the effective theory
that stem from it are trivial to construct because they are chirally invariant. Those with $w_Q - \ol w_Q$
have the flavor symmetry of the mixed action. Let us write out one such term from $\cL^{(6)}$, 
\begin{equation}
\ol Q \Dslash {}^3 (w_Q - \ol w_Q) Q
=
\ol Q_L \Dslash {}^3 (w_Q - \ol w_Q) Q_L
+ 
\ol Q_R \Dslash {}^3 (w_Q - \ol w_Q) Q_R 
\end{equation}
To make these terms $SU(4|2)$ chirally invariant, we introduce two spurions $\cO_1$ and $\cO_2$ transforming as
\begin{eqnarray}
\cO_1 &\to& L \cO_1 L^\dagger \notag \\
\cO_2 &\to& R \cO_2 R^\dagger,  
\end{eqnarray}
that will be assigned the values $\cO_1 = \cO_2 = w_Q - \ol w_Q$. 
In the effective field theory, the operators $\cO = \xi^\dagger \cO_{1} \xi$, and $\xi \cO_2 \xi^\dagger$ both  
transform as $U \cO U^\dagger$. Instead of using these operators, we work with 
linear combinations of definite parity, namely
\begin{equation}
\cO_\pm = \frac{1}{2} 
\left[ 
\xi^\dagger (w_Q - \ol w_Q ) \xi \pm \xi (w_Q - \ol w_Q) \xi^\dagger
\right]
.\end{equation}
Thus class one operators get mapped into the effective theory as terms contained in the Lagrangian
\begin{eqnarray} \label{eq:A2s}
\cL &=& - a^2 \L_{QCD}^3 \Bigg[ 
b_0  \left( \ol \cB \cB \right) 
+
b_1^\cO  \left( \ol \cB \cB \cO_+ \right)
+ 
b_2^\cO \left( \ol \cB \cO_+ \cB \right)
+ 
b_3^\cO \left( \ol \cB \cB \right) \str \left( \cO_+ \right) \notag \\
&&
-
t_0 \left( \ol \cT_\mu \cT_\mu \right) 
- 
t_1^\cO \left( \ol \cT_\mu \cO_+ \cT_\mu \right) 
- 
t_2^\cO \left( \ol \cT_\mu \cT_\mu \right) \str \left( \cO_+ \right)
\Bigg].
\end{eqnarray}
When the action is unmixed, the operator $\cO_+$ is proportional to the identity matrix and hence
there is only one independent operator for the $\cB$-field and one for the $\cT^\mu$-field 
in Eq.~\eqref{eq:A2s}.

All class two operators have an insertion of the quark mass matrix $m_Q$. Thus these operators are at least 
of $\cO( m_Q \, a^2) = \cO(\e^6)$ in our power counting and can be neglected to the order we are working. 
Class three operators are four-quark operators that do not break chiral symmetry in the valence and sea sectors.
They do break the $SU(4|2)$ chiral symmetry down to the chiral symmetry $G$ of the mixed action.
The requisite spurions for these class three operators are $\cO_1 \otimes \cO_1$, $\cO_1 \otimes \cO_2$, $\cO_2 \otimes \cO_1$, 
and $\cO_2 \otimes \cO_2$. Hence operators in the effective theory will involve the products $\cO_+ \otimes \cO_+$ and
$\cO_- \otimes \cO_-$. The latter do not contribute to baryon masses to the order we work, while the former
baryon operators are contained in the Lagrangian
\begin{eqnarray}
\cL &=&
- a^2 \L_{QCD}^3 \Bigg[
b_1^{\cO\cO} (-)^{(\eta_i + \eta_j)(\eta_k + \eta_{k'})} \ol \cB {}^{kji} \left( \cO_+ \cO_+ \right)^{kk'} \cB^{ijk'}
+
b_2^{\cO\cO} \ol \cB {}^{kji} \left( \cO_+ \cO_+ \right)^{ii'} \cB^{i'jk}
\notag \\
&& + 
b_3^{\cO\cO} (-)^{\eta_{i'} ( \eta_j + \eta_{j'})} \ol \cB {}^{kji} \cO_+^{ii'} \cO_+^{jj'} \cB^{i'j'k}
+ 
b_4^{\cO\cO} (-)^{\eta_j \eta_{j'} +1} \ol \cB {}^{kji} \cO_+^{ij'} \cO_+^{ji'} \cB^{i'j'k}
\notag \\
&& + 
b_5^{\cO\cO} \left( \ol \cB \cB  \cO_+ \right) \str (\cO_+)
+ 
b_6^{\cO\cO} \left( \ol \cB \cO_+ \cB \right) \str (\cO_+) 
+
b_7^{\cO\cO} \left( \ol \cB \cB \right) \str (\cO_+ \cO_+) 
\notag \\
&& + 
b_8^{\cO\cO} \left( \ol \cB \cB \right) \str (\cO_+) \str (\cO_+) 
\Bigg] 
,\label{eq:OOBs}\end{eqnarray}
for the spin-$\frac{1}{2}$ fields, and 
\begin{eqnarray}
\cL &=&
a^2 \L_{QCD}^3 \Bigg[
t_1^{\cO\cO} \ol \cT {}^{kji}_\mu \left( \cO_+ \cO_+ \right)^{ii'} \cT^{i'jk}_\mu
+
t_2^{\cO\cO} (-)^{\eta_{i'} ( \eta_j + \eta_{j'})} \ol \cT {}^{kji}_\mu \cO_+^{ii'} \cO_+^{jj'} \cT^{i'j'k}_\mu
\notag \\
&& + 
t_3^{\cO\cO} \left( \ol \cT_\mu \cO_+ \cT_\mu \right) \str (\cO_+) 
+ 
t_4^{\cO\cO} \left( \ol \cT_\mu \cT_\mu \right) \str (\cO_+ \cO_+)
+ 
t_5^{\cO\cO} \left( \ol \cT_\mu \cT_\mu \right) \str (\cO_+ ) \str (\cO_+)
\Bigg],
\notag \\
\label{eq:OOTs}
\end{eqnarray}
for the spin-$\frac{3}{2}$ fields. 
When considering unmixed actions, all of the operators in Eqs.~\eqref{eq:OOBs} and \eqref{eq:OOTs} become 
redundant compared to those in Eq.~\eqref{eq:A2s}, because $\cO_+$ is proportional to the identity.

Class four operators are four-quark operators that break chiral symmetry in the Wilson sector of the theory.
Such terms involve the flavor structure $w_Q \otimes w_Q$, which must be promoted to various spurions.  
In the Symanzik Lagrangian, for example, we have the operator
\begin{eqnarray}
\left( \ol Q w_Q Q \right)^2 
&=&  
\left( \ol Q_L w_Q Q_R \right)^2 
+  
\left( \ol Q_L w_Q Q_R \right) \left( \ol Q_R w_Q Q_L \right) \notag \\
&& + 
\left( \ol Q_R w_Q Q_L \right) \left( \ol Q_L w_Q Q_R \right)
+
\left( \ol Q_R w_Q Q_L \right)^2
,\end{eqnarray}
and so we require the spurions~\cite{Bar:2003mh}
\begin{eqnarray}
B_1 \otimes B_2 &\to& L B_1 R^\dagger \otimes L B_2 R^\dagger \notag \\
B_1^\dagger \otimes B_2^\dagger &\to& R B_1^\dagger L^\dagger \otimes R B_2^\dagger L^\dagger \notag \\
C_1 \otimes C_2 &\to& R C_1 L^\dagger \otimes L C_2 R^\dagger \notag \\
C_1^\dagger \otimes C_2^\dagger &\to& L C_1^\dagger R^\dagger \otimes R C_2^\dagger L^\dagger, 
\end{eqnarray}
that will ultimately be given the values $B_1 = B_2 = C_1 = C_2 = a \L_{QCD}^2 w_Q$, and similarly for their
Hermitian conjugates. Now the operators $\cO = \xi^\dagger B_{1,2} \xi^\dagger$, $\xi B_{1,2}^\dagger \xi$, 
$\xi C_1 \xi$, $\xi^\dagger C_1^\dagger \xi^\dagger$, $\xi^\dagger C_2 \xi^\dagger$, and
$\xi C_2^\dagger \xi$ all transform as $U \cO U^\dagger$  under their respective spurion transformations. 
Moreover when assigned constant values for their spurions, the operators involving $C$'s and $B_2$'s become indistinguishable from those 
involving $B_1$ and $B_1^\dagger$. 
Finally instead of working with the operators $\xi w_Q \xi$ and $\xi^\dagger w_Q \xi^\dagger$, we work with linear 
combinations that are parity even and odd, which are the $\cW_\pm$ operators, respectively, that were introduced previously.  
Thus our spurion analysis shows  
terms in the effective theory will involve products of Wilson operators 
$\cW_+ \otimes \cW_+$, and  $\cW_- \otimes \cW_-$.\footnote{%
As spurions, these have the same effect as squares of the $\cO(a)$ spurions. Thus
we need not consider such higher-order effects from lower order terms.
} 
The latter products do not contribute to the baryon masses at this order. 
Thus for the spin-$\frac{1}{2}$ baryons, we have the eight terms from class 
four operators
\begin{eqnarray}
\cL &=&
- \frac{1}{\L_{QCD}} \Bigg[
b_1^W (-)^{(\eta_i + \eta_j)(\eta_k + \eta_{k'})} \ol \cB {}^{kji} \left( \cW_+ \cW_+ \right)^{kk'} \cB^{ijk'}
+
b_2^W \ol \cB {}^{kji} \left( \cW_+ \cW_+ \right)^{ii'} \cB^{i'jk}
\notag \\
&& + 
b_3^W (-)^{\eta_{i'} ( \eta_j + \eta_{j'})} \ol \cB {}^{kji} \cW_+^{ii'} \cW_+^{jj'} \cB^{i'j'k}
+ 
b_4^W (-)^{\eta_j \eta_{j'} +1} \ol \cB {}^{kji} \cW_+^{ij'} \cW_+^{ji'} \cB^{i'j'k}
\notag \\
&& + 
b_5^W \left( \ol \cB \cB  \cW_+ \right) \str (\cW_+)
+ 
b_6^W \left( \ol \cB \cW_+ \cB \right) \str (\cW_+) 
+
b_7^W \left( \ol \cB \cB \right) \str (\cW_+ \cW_+) 
\notag \\
&& + 
b_8^W \left( \ol \cB \cB \right) \str (\cW_+) \str (\cW_+) 
\Bigg] 
;\label{eq:WBs}\end{eqnarray}
while in the spin-$\frac{3}{2}$ sector, we have five terms
\begin{eqnarray}
\cL &=&
\frac{1}{\L_{QCD}} \Bigg[
t_1^W \ol \cT {}^{kji}_\mu \left( \cW_+ \cW_+ \right)^{ii'} \cT^{i'jk}_\mu
+
t_2^W (-)^{\eta_{i'} ( \eta_j + \eta_{j'})} \ol \cT {}^{kji}_\mu \cW_+^{ii'} \cW_+^{jj'} \cT^{i'j'k}_\mu
\notag \\
&& + 
t_3^W \left( \ol \cT_\mu \cW_+ \cT_\mu \right) \str (\cW_+) 
+ 
t_4^W \left( \ol \cT_\mu \cT_\mu \right) \str (\cW_+ \cW_+)
+ 
t_5^W \left( \ol \cT_\mu \cT_\mu \right) \str (\cW_+ ) \str (\cW_+)
\Bigg].
\notag \\
\label{eq:WTs}
\end{eqnarray}

Class five operators break the $O(4)$ rotational symmetry of Euclidean space. 
The lowest-order hypercubic invariants that are parity even are: $v_\mu v_\mu v_\mu v_\mu$, 
$v_\mu v_\mu S_\mu S_\mu$,  and $S_\mu S_\mu S_\mu S_\mu$. The latter two are both redundant
because, suspending the Einstein summation convention, we have 
$S_\mu S_\mu = \frac{1}{4} (\delta_{\mu \mu} + v_\mu v_\mu )$.
There is an additional hypercubic invariant for the spin-$\frac{3}{2}$ fields since they 
carry a vector index. 
Furthermore, these class five operators have the chiral symmetry of the group $G$ of the mixed action, not 
the full graded chiral symmetry of $\cL^{(4)}$. 
As with class one operators, class five operators in the effective theory 
require the insertion of the operator $\cO_+$. Thus the $O(4)$ breaking operators are
\begin{eqnarray}
\cL 
&=& 
- a^2 \L_{QCD}^3 \Bigg[
b^v_0 \, \left( \ol \cB  v_\mu v_\mu v_\mu v_\mu \cB \right)
+
b^v_1 \, \left( \ol \cB  v_\mu v_\mu v_\mu v_\mu \cB \cO_+\right)
+
b_2^v \, \left( \ol \cB  v_\mu v_\mu v_\mu v_\mu \cO_+ \cB \right)
\notag \\
&& + 
b_3^v \, \left( \ol \cB  v_\mu v_\mu v_\mu v_\mu \cB \right) \str \left( \cO_+ \right) 
-
t_0^{v} \, \left(  \ol \cT_\nu v_\mu v_\mu v_\mu v_\mu \cT_\nu \right)
-
t_1^{v} \, \left(  \ol \cT_\nu v_\mu v_\mu v_\mu v_\mu \cO_+ \cT_\nu \right)
\notag \\
&& 
-
t_2^v \, \left(  \ol \cT_\nu v_\mu v_\mu v_\mu v_\mu \cT_\nu \right) \str \left( \cO_+ \right)
- 
t_0^{\ol v} \,  \left( \ol \cT_\mu v_\mu v_\mu \cT_\mu \right)
- 
t_1^{\ol v} \,  \left( \ol \cT_\mu v_\mu v_\mu \cO_+ \cT_\mu \right)
\notag \\
&& 
- 
t_2^{\ol v} \,  \left( \ol \cT_\mu v_\mu v_\mu \cT_\mu \right) \str \left( \cO_+ \right)
\Bigg].
\label{eq:vA2s}
\end{eqnarray}

The mixing of classes of operators is only possible through linear superpositions. This 
is accounted for in the effective theory by linear superpositions of effective operators 
and thus need not be addressed separately. 
This completes the (nearly) exhaustive listing of $\cO(\e^4)$
\PQCPT\ terms for the mixed lattice action. 
Despite the large number of operators (over one hundred), many terms do not contribute
to baryon masses at tree level and have already been omitted. 
Additionally many of the contributing terms are not linearly 
independent at leading order. 
In the following section we calculate the masses of the nucleons and deltas to $\cO(a^2)$,
and thereby determine the number of new free parameters that are required.

\section{Baryon masses to $\cO(a^2)$} \label{s:mass}

In this section we calculate the masses of the nucleons and deltas in $SU(4|2)$. 
Baryon masses in $SU(2)$ are presented in Appendix~\ref{s:2}.

\subsection{Nucleon mass} \label{s:nm}

In Euclidean lattice field theory, the baryon self energy $\Sigma(\gamma_\mu p_\mu)$ is 
no longer rotationally invariant; it is a hypercubic invariant function of the baryon momentum and 
gamma matrices. With this functional dependence in mind, we can now calculate the mass of the nucleon. 
The nucleon mass in the combined lattice spacing and chiral expansion can be written as
\begin{eqnarray}
     M_{N} = M_0 \left(\mu \right) -  M_{N}^{(2)}\left(\mu \right)
                - M_{N}^{(3)}\left(\mu \right)
                - M_{N}^{(4)}\left(\mu \right) + \ldots
\label{eq:Bmassexp}
\end{eqnarray}
Here, $M_0 \left(\mu \right)$ is the renormalized nucleon
mass in the continuum and chiral limits, 
which is independent of $m_Q$.
$M_{N}^{(n)}$ is the contribution to the mass of order $\e^{n}$, and $\mu$ is the
renormalization scale.

At order $\e^2$ in the combined lattice spacing and chiral expansion, we have 
contributions at tree level from the $\cO(m_Q)$ and $\cO(a)$ operators in Eq.~\eqref{eqn:L}.
These contributions to the nucleon mass read
\begin{equation}
M_N^{(2)} 
= 
2 (\a_M  + \b_M ) m_u   + 4  \sigma_M \, m_j  
+
2 a \L_{QCD}^2 
\left[
( \a_W + \b_W) w_v 
+ 2 \sigma_W \, w_s 
\right] \label{eq:Bmass2}
.\end{equation}
From this expression, we see that the $\cO(a)$ mass corrections vanish only if
both the valence and sea quarks are GW. This is in contrast to the meson sector
where only the valence quarks need be GW to ensure the vanishing of linear $a$ contributions.
At $\cO(\e^3)$ there are contributions from loop graphs to $M^{(3)}_N$. 
These have been determined for $SU(4|2)$ in~\cite{Beane:2002vq} and we 
do not duplicate the expressions here because the only modification necessary
is to include the $a$-dependence of the loop meson masses via Eq.~\eqref{eqn:mqq}.\footnote{% 
Additionally there are local operators that contribute at this order. The form of these operators, 
however, is the same as those in Eq.~\eqref{eqn:L} but multiplied by a factor of $\D / \L_\chi$. 
We can thus trivially include the effect of these operators by promoting the 
LECs to arbitrary polynomial functions of $\D / \L_\chi$ expanded out to the appropriate order.  
We do not spell this out explicitly since the determination of these polynomial coefficients requires
the ability to vary $\D$. }

\begin{figure}
\epsfig{file=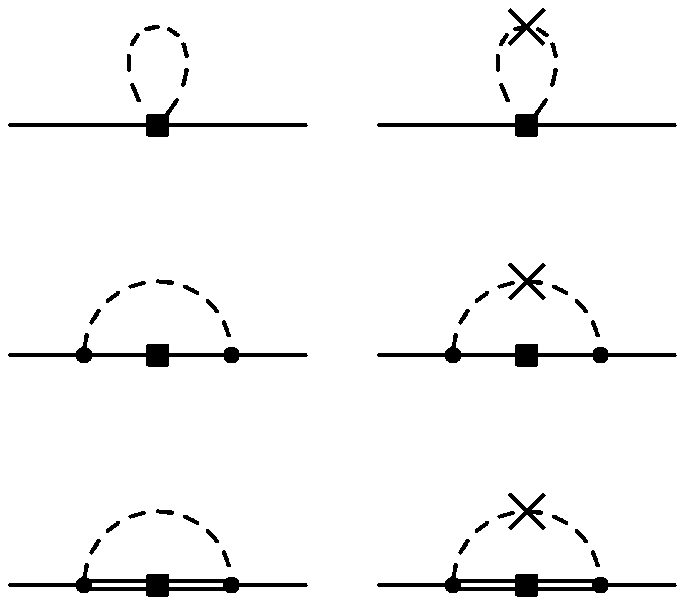}
\caption{Loop diagrams contributing to the $a$-dependence of the nucleon mass at $\cO(\e^4)$. 
Mesons are denoted by a dashed line, flavor neutrals (hairpins) by a crossed dashed line, 
and a thin solid line denotes an octet baryon. The square denotes an $\cO(a)$ vertex. 
}
\label{F:amass}
\end{figure}

At $\cO(\e^4)$, we have the usual continuum contributions from tree-level $\cO(m_Q^2)$ operators and from 
loop diagrams. These have been detailed for $SU(4|2)$ in \cite{Tiburzi:2005na} and we do not duplicate these
lengthy expressions here. The only modification necessary at finite lattice spacing is the inclusion 
of the $a$-dependence of the loop meson masses from Eq.~\eqref{eqn:mqq}.  The remaining finite lattice spacing corrections
arise at tree level from the $\cO(m_Q \, a)$ operators in Eq.~\eqref{eq:BAMs} and the $\cO(a^2)$ operators in Eqs.~\eqref{eq:A2s}, 
\eqref{eq:WBs}, and \eqref{eq:vA2s}. 
Additionally there are loop diagrams arising from the operators in Eq.~\eqref{eqn:L}, where
either the Wilson operator in Eq.~\eqref{eq:Wilson} is expanded to second order or is inserted
on an internal baryon line. These diagrams are shown in Fig.~\ref{F:amass}. Lastly there are wavefunction
renormalization corrections that are linear in $a$. 
The contribution to the nucleon mass from these operators and loops appears in $M_N^{(4)}$ as a correction of the form
\begin{eqnarray}
\delta M_N^{(4)} 
&=& 
\frac{a \L_{QCD}^2}{\L_\chi} 
\Big[
A  w_v  m_u
+ 
B  w_s m_u
+ 
C  w_v  m_j
+ 
D  w_s  m_j
\Big]
\notag \\
&& - 4 \frac{a \L_{QCD}^2}{\L_\chi^2}
\Bigg\{
(\a_W + \b_W)  
\left[ 
(w_v + w_s)
\cL(m_{ju}, \mu)
+ 
w_v
\cL(m_{\eta_u}, m_{\eta_u}, \mu)
\right]
\notag \\
&& \phantom{spacer} + 
2 \sigma_W \, w_s
\left[ 
2 \cL(m_{jj}, \mu)
+ 
\cL(m_{\eta_j}, m_{\eta_j}, \mu)
\right]
+
A^N_{ju} \left[ \cL(m_{ju}, \mu) + \frac{2}{3} m_{ju}^2 \right]
\notag \\
&& \phantom{spacer} + 
g_{\D N}^2 [B^N_\pi \cJ(m_\pi, \D, \mu) + B^N_{ju} \cJ(m_{ju}, \D, \mu) ]
\Bigg\}
\notag \\
&& +
a^2 \L_{QCD}^3 
\Big[E+ E' w_v +E'' w_s+E''' w_v w_s
\notag \\
&& \phantom{spacer}
+ 
\ol u(p) v_\mu v_\mu v_\mu v_\mu \, u(p)
( F + F' w_v + F'' w_s )
\Big] \label{eq:Bmass4}
.\end{eqnarray}
The non-analytic functions appearing in this expression are defined by
\begin{eqnarray}
\cL (m_\phi, \mu) &=& m_\phi^2 \log \frac{m_\phi^2}{\mu^2}, \\
\cJ ( m_\phi, \delta, \mu ) &=&
(m_\phi^2 - \d^2) \log \frac{m_\phi^2}{\mu^2} + 2 \d \sqrt{\d^2 - m_\phi^2} \log \left( \frac{\d - \sqrt{\d^2 - m_\phi^2 + i \epsilon}}
{\d + \sqrt{\d^2 - m_\phi^2 + i \epsilon}} \right), \\
%\cM^2(m_\phi, m_{\phi'}) &=& \cH_{\phi \phi'} \left[ m_\phi^2, m_{\phi'}^2 \right] \\
\cL(m_\phi, m_{\phi'}, \mu) &=& \cH_{\phi \phi'} \left[ \cL(m_\phi,\mu), \cL(m_{\phi'},\mu) \right], \\
\cJ(m_\phi, m_{\phi'},\d,\mu) &=& \cH_{\phi \phi'} \left[ \cJ(m_\phi,\d, \mu), \cJ(m_{\phi'},\d, \mu) \right] 
,\end{eqnarray}
and one should keep in mind the $a$-dependence of the meson masses. 
The parameters $A$--$F$ are 
replacements for particular combinations of LECs.\footnote{%
For clarity in comparing with the operators written down in Sec.~\ref{s:baryons}, these combinations are:
$A = 2 b_1^{WM} + 2 b_2^{WM} +  b_3^{WM}+  b_4^{WM} + 2 b_5^{WM}$, 
$B = 2 b_6^{WM} + 2 b_7^{WM}$, 
$C = 2 b_{10}^{WM} + 2  b_{11}^{WM}$, 
$D = 2 b_8^{WM} + 2 b_9^{WM}$, 
$E = b_0 - b_1^\cO - b_2^\cO - 2 b_3^\cO + b_1^{\cO \cO} + b_2^{\cO \cO} + b_3^{\cO \cO} + b_4^{\cO \cO} + 2 b_5^{\cO \cO} 
+ 2 b_6^{\cO \cO} + 2 b_7^{\cO \cO} + 4 b_8^{\cO \cO}$, 
$E' = b_1^W + b_2^W + b_3^W + b_4^W + 2 b_1^\cO + 2 b_2^\cO - 4 b_5^{\cO \cO} - 4 b_6^{\cO \cO}$, 
$E'' = 2 b_7^W + 4 b_8^W + 4 b_3^\cO - 4 b_5^{\cO \cO} - 4 b_6^{\cO \cO}$, 
$E''' = 2 b_5^W + 2 b_6^W + 8 b_5^{\cO \cO} + 8 b_6^{\cO \cO}$, 
$F = b^v_0 - b_1^v - b_2^v - 2 b_3^v$,
$F' = 2 b_1^v + 2 b_2^v$,
and
$F'' = 4 b_3^v$.  
}
The meson loop coefficients $A^N_{ju}$, $B^N_\pi$, and $B^N_{ju}$ are given by
\begin{eqnarray}
A^N_{ju} &=& -\frac{1}{4} \Bigg\{ 2 w_v \left[ - 2 g_A^2 \a_W + g_A g_1 \b_W - \frac{1}{2} g_1^2 (\a_W + 3 \b_W) \right] 
\notag \\
&& \phantom{spac}
+ w_s \left[ ( 2 g_A^2 + g_A g_1) \a_W + g_1^2 ( \a_W + 3 \b_W)  \right] \Bigg\}, \notag \\
B^N_\pi &=& w_v ( \a_W + \b_W + \gamma_W ) + 2 w_s ( \sigma_W - \ol \sigma_W ), \notag \\
B^N_{ju} &=& w_v \left( \a_W + \b_W + \frac{2}{3} \gamma_W \right) + w_s \left( 2 \sigma_W - 2 \ol \sigma_W + \frac{1}{3} \gamma_W \right) 
.\end{eqnarray}

Despite the complicated form of Eq.~\eqref{eq:Bmass4}, there are at most only four free parameters depending 
on the mixed action considered. The reason for the complicated form is that the above equation
shows the interplay between all possible theories with mixed actions.  Precisely which operators 
are present or, on the other hand, can be eliminated is perhaps of academic interest because
at this order many of the individual contributions cannot be resolved from lattice data. 
For example, in a theory with Wilson valence and sea quarks, the contribution $A + B$ cannot be further resolved. 
In Table~\ref{t:summary}, we
summarize the number of free parameters introduced by $\cO(a)$, $\cO(m_Q \,a)$, 
and $\cO(a^2)$ operators in the baryon sector for various partially quenched theories. 
Of course, the simplest situation occurs in a theory with GW valence and sea quarks. Here
we see that there are $\cO(a^2)$ corrections to the nucleon self energy of two forms, one
of which violates $O(4)$ symmetry.

Let us investigate the violation of $O(4)$ rotational invariance in detail. For an on-shell nucleon
at rest, the $O(4)$ breaking operator merely contributes an additive $a^2$ shift to the mass. 
This correction with $F$ coefficients becomes indistinguishable from the $a^2$ corrections 
which respect $O(4)$ [these have $E$ coefficients in Eq.~\eqref{eq:Bmass4}]. 
For any particle with non-relativistic three-momentum $\bm{p}$, 
however, we can see the effects of $O(4)$ violation in the dispersion relation.
To write the dispersion relation compactly, let us assume both valence and sea quarks are GW.
The same form of the dispersion relation holds for actions involving Wilson quarks with the addition
of various $\cO(a)$ contributions. First let us ignore the contributions from $O(4)$ breaking operators. 
The dispersion relation has the familiar form
\begin{equation} \label{eq:familiar}
E_{\bm{p}} = M(a) + \frac{\bm{p}^2}{2 M(a)} + \ldots
,\end{equation}
where $M(a) = M(0) - a^2 \L_{QCD}^2 E$, and $M(0)$ denotes the particle mass in the continuum 
limit (which remains quark mass dependent).

To add the $O(4)$ breaking terms to the dispersion relation, we can use a 
non-relativistic expansion for the particle four-velocity or utilize reparameterization 
invariance to deduce the corrections to the dispersion relation. We relegate this 
discussion to Appendix~\ref{s:RPI}, where we address particles with spin less than two. 
Using the result in the Appendix for the spin-$1/2$ nucleon, 
the relation between the particle energy and momentum is then given by
\begin{eqnarray} \label{eq:dispersion}
E_{\bm{p}} 
&=& 
M(0) 
-
a^2 \L_{QCD}^3 (E + F) 
+ 
\frac{\bm{p}^2}{2 M(0)} \left[ 1 + \frac{a^2 \L_{QCD}^3}{M(0)} ( E + 5 F ) \right] 
.\end{eqnarray}
Here we have written the non-relativistic expansion to second order and have retained the leading lattice-spacing 
correction for each term. Notice for a particle at rest, we recover the simple result of an $a^2$ shift. 
The kinetic energy receives corrections at $\cO(a^2)$, these are due to the $a^2$ shift in mass as well 
as the $O(4)$ contribution in Eq.~\eqref{eq:vvvv}. If it were not for the latter contribution, the dispersion
relation would retain the form in Eq.~\eqref{eq:familiar} with $M(a) = M(0) - a^2 \L_{QCD}^2 ( E + F)$. 
Additionally the relativistic correction to the kinetic energy will be modified with a coefficient 
proportional to $a^2$ and a different linear combination of $E$ and $F$. Also the energy 
is only a cubic invariant function of the spatial momenta $p_i$. Because four factors of the momentum are required, 
the leading $O(3)$ breaking term in the energy occurs at $\cO(a^2 M^{-4})$ with a coefficient $- F$.

\subsection{Delta mass} \label{s:dm}

For the masses of spin-$\frac{3}{2}$ baryons, there is an additional feature on a hypercube due to the fact
that the Rarita-Schwinger fields themselves carry a vector index. 
The one-particle irreducible self energy for a spin-$\frac{3}{2}$ field on a hypercube has the form
\begin{equation}
\ol u_\mu(p) \, \Sigma_{\mu \nu} \, u_\nu (p) 
= 
\ol u_\mu(p) \, 
\left( 
\Sigma + \ol \Sigma_{\mu \mu} 
\right) u_\mu (p) 
,\end{equation}
where $\Sigma$ and $\ol \Sigma_{\mu\mu}$ are both hypercubic invariant functions. 
With this in mind, we can now calculate $\cO(a)$, $\cO(m_Q \, a)$ and $\cO(a^2)$ 
contributions to the delta mass.
The mass of the delta in the combined lattice spacing and chiral expansion can be written as
\begin{eqnarray}
     M_{\D} = M_0 \left(\mu \right) + \D(\mu) +  M_{\D}^{(2)}\left(\mu \right)
                + M_{\D}^{(3)}\left(\mu \right)
                + M_{\D}^{(4)}\left(\mu \right) + \ldots
\label{eq:Tmassexp}
\end{eqnarray}
Here, $M_0 \left(\mu \right)$ is the renormalized baryon mass 
in the continuum and chiral limits and $\D(\mu)$ is the renormalized mass splitting between the spin-$\frac{3}{2}$ and spin-$\frac{1}{2}$
baryons in the continuum and chiral limits. Both of these quantities are independent of $a$ and $m_Q$.
$M_{\D}^{(n)}$ is the contribution of order $\e^{n}$, and $\mu$ is the
renormalization scale.

At order $\e^2$ in the expansion, we have contributions to the delta mass 
at tree level from the $\cO(m_Q)$ and $\cO(a)$ operators in Eq.~\eqref{eqn:L}.
These lead to
\begin{equation}
M_\D^{(2)} 
= 
2 \gamma_M  \, m_u 
- 
4 \ol \sigma_M \, m_j
+ 
2 a \L_{QCD}^2
\left[
\gamma_W \, w_v
- 
2 \ol \sigma_W \, w_s
\right] \label{eq:Tmass2}
.\end{equation}
Again in contrast with the mesons,  we note that both valence and sea quarks must be GW for the $\cO(a)$ delta mass corrections to vanish.
At $\cO(\e^3)$ there are contributions from loop graphs to $M^{(3)}_\D$. 
These have been determined for $SU(4|2)$ in~\cite{Tiburzi:2005na} and we 
do not duplicate the expressions here, because the only modification necessary
is to include the $a$-dependence of the loop meson masses via Eq.~\eqref{eqn:mqq}.

\begin{figure}
\epsfig{file=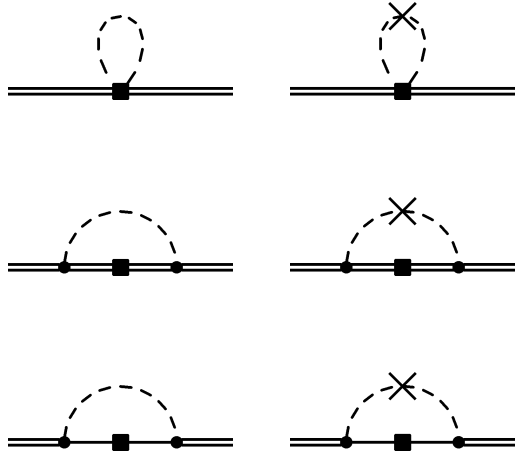}
\caption{Loop diagrams contributing to the $a$-dependence of the delta mass. 
Mesons are denoted by a dashed line, flavor neutrals (hairpins) by a crossed dashed line, 
and a thick solid line denotes an decuplet baryon. The square denotes an $\cO(a)$ vertex. 
}
\label{F:decmass}
\end{figure}

Finally at $\cO(\e^4)$ there are the usual contributions from tree-level $\cO(m_Q^2)$ operators and from 
loop diagrams. For the delta, these have been detailed for $SU(4|2)$ in \cite{Tiburzi:2005na}.
The remaining finite lattice-spacing corrections
arise at tree level from the $\cO(m_Q \, a)$ operators in Eq.~\eqref{eq:TAMs} and the $\cO(a^2)$ operators in Eqs.~\eqref{eq:A2s}, 
\eqref{eq:WTs}, and \eqref{eq:vA2s}. Additionally there are loop contributions stemming from the 
operators in Eq.~\eqref{eqn:L}. These loop diagrams are depicted in Fig.~\ref{F:decmass}. Lastly 
we must include wavefunction renormalization corrections that are linear in $a$.  
The contribution to the delta mass from these operators and loops yields
\begin{eqnarray}
\delta M_\D^{(4)} 
&=& 
\frac{a \L_{QCD}^2}{\L_\chi} \Big[
A  w_v   m_u
+ 
B  w_s m_u
+ 
C w_v m_j
+ 
D w_s m_j
\Big]
\notag \\
&& 
- 4 \frac{a \L_{QCD}^2}{\L_\chi^2} 
\Bigg\{
\g_W 
\left[ 
(w_v + w_s) \cL(m_{ju}, \mu)
+ 
w_v
\cL(m_{\eta_u}, m_{\eta_u}, \mu)
\right]
\notag \\
&& \phantom{spacs} \, - 
2 \ol \sigma_W \, w_s
\left[ 
2 \cL(m_{jj}, \mu)
+ 
\cL(m_{\eta_j}, m_{\eta_j}, \mu)
\right]
\notag \\
&& \phantom{spacs}
+ \frac{5}{27} \gamma_W g_{\D \D}^2 (w_v - w_s) \left[ \cL (m_{ju},\mu) + \frac{26}{15} m_{ju^2} \right]
\notag \\
&& \phantom{spacs}
- 
\frac{1}{2}  g_{\D N}^2 \left[ B^\D_\pi \cJ(m_\pi, -\D, \mu) + B^\D_{ju} \cJ(m_{ju}, - \D, \mu) \right]
\Bigg\}
\notag \\
&& +
a^2 \L_{QCD}^3 
\Big[
E 
+ 
E'  w_v 
+
E'' w_s
+ 
E''' w_v w_s
\notag \\
&& \phantom{spacs}
+ 
\ol u(p)_\nu v_\mu v_\mu v_\mu v_\mu \, u_\nu(p)
\Big(
F + F' w_v + F'' w_s
\Big)
\notag \\
&& \phantom{spacs}
+ 
\ol u_\mu (p) v_\mu v_\mu \, u_\mu(p) 
\Big( 
G + G' w_v + G'' w_s
\Big) 
 \Big] \label{eq:Tmass4}
.\end{eqnarray}
The meson loop coefficients $B^\D_\pi$ and $B^\D_{ju}$ are given by
\begin{eqnarray}
B^\D_\pi &=& w_v ( \a_W + \b_W + \gamma_W ) + 2 w_s ( \sigma_W - \ol \sigma_W ),  \notag \\
B^\D_{ju} &=& - \frac{1}{3} [w_v ( 5 \a_W + 2 \b_W + 6 \gamma_W ) + w_s (\a_W + 4 \b_W + 12 \sigma_W - 12 \ol \sigma_W)] \notag
.\end{eqnarray} 
The complicated nature of this expression results from considering general theories with mixed actions. 
The formula retains book-keeping factors that indicate which operators contribute in particular mixed action theories. 
Notice, for example, that for a theory with Wilson valence and sea quarks, the combination $A + B$ 
cannot be further resolved from lattice data. Here the parameters $A$--$G$ depend on particular linear combinations
of the LECs in the spin-$\frac{3}{2}$ sector.\footnote{%
For clarity in comparing with Sec.~\ref{s:baryons}, these combinations of LECs are:
$A = 2 t_1^{WM} +  t_2^{WM}$, 
$B = 2 t_3^{WM}$, 
$C = 2 t_6^{WM}$, 
$D = 2 t_4^{WM} + 4 t_5^{WM}$, 
$E = t_0 - t_1^\cO - 2 t_2^\cO + t_1^{\cO \cO} + t_2^{\cO \cO} +2 t_3^{\cO \cO} + 2 t_4^{\cO \cO} + 4 t_5^{\cO \cO}$, 
$E' = t_1^W + t_2^W + 2 t_1^\cO - 4 t_3^{\cO \cO}$, 
$E'' = 2 t_4^W + 4 t_5^W+ 4 t_2^\cO - 4 t_3^{\cO \cO}$, 
$E''' =2 t_3^W + 8 t_3^{\cO \cO}$, 
$F = t^{v}_0 - t^{v}_1 - 2 t^{v}_2$,
$F' = 2 t^{v}_1$, 
$F'' = 4 t^{v}_2$.
$G = t^{\ol v}_0 - t^{\ol v}_1 - 2 t^{\ol v}_2$, 
$G' = 2 t^{\ol v}_1$,
and
$G'' = 4 t^{\ol v}_2$,
}
The number of free parameters introduced for Wilson valence and sea quarks from
$\cO(a)$, $\cO(m_Q\, a)$, and $\cO(a^2)$ operators is six.
Table~\ref{t:summary} lists the number of free parameters
introduced from finite lattice spacing effects for delta mass in mixed action theories. 
The simplest expression results from an unmixed action of GW quarks. Corrections of 
course start at $\cO(a^2)$, but are of three distinct forms given the possible hypercubic invariants.

As we know from determining the nucleon mass above, the $O(4)$ violating terms proportional to 
$v_\mu v_\mu v_\mu v_\mu$ in Eq.~\eqref{eq:Tmass4} with coefficients $F$ give rise to corrections
to the dispersion relation. There are additional corrections with coefficients $G$ that are tied 
to the Rarita-Schwinger spinors in the form $\ol u_\mu(p) v_\mu v_\mu \, u_\mu(p)$. 
Because of the constraint $v_\mu \, u_\mu(p) = 0$, these new terms
do not contribute for a particle at rest. To consider a particle moving with non-relativistic three momentum $\bm{p}$, 
we must use explicit forms for the Rarita-Schwinger spinors to obtain the modification to the energy-momentum 
relation, see Appendix~\ref{s:RPI}. For GW valence and sea quarks, we find a similar dispersion relation
given in Eq.~\eqref{eq:dispersion} but with an additional term 
\begin{eqnarray} \label{eq:new}
E^\l_{\bm{p}} &=&
M(0) 
+
a^2 \L_{QCD}^3 (E + F) 
+ 
\frac{\bm{p}^2}{2 M(0)} \left[ 1 - \frac{a^2 \L_{QCD}^3}{M(0)} ( E - 3 F ) \right] 
\notag \\ 
&& +
\frac{a^2 \L_{QCD}^3 G}{3 M(0)^2} 
\left[
|\l| \bm{p}_\perp^2 - \left( 2 |\l| - 3 \right) \bm{p}_\l^2
\right]
,\end{eqnarray} 
where $\lambda = \pm 3/2, \pm 1/2$ refer to the delta spin states quantized with respect to an axis,
$\bm{p}_\l$ is the momentum along that axis, 
and $\bm{p}_\perp$ is the momentum transverse to that axis. 
We have retained only the leading term
in the non-relativistic expansion and its $a^2$ correction. Unlike the nucleon, the delta
dispersion relation is more sensitive to $O(3)$ violation which enters at $\cO(a^2 M^{-2})$ as opposed
to $\cO(a^2 M^{-4})$. The splitting of spin states is proportional to $|\l|$ because one has 
invariance under rotations of the quantization axis by $\pi$. For an unpolarized state, however,
the $O(3)$ violation in Eq.~\eqref{eq:new} averages out.

\begin{table}
\caption{Summary of the new free parameters entering particle masses in chiral perturbation theory
for non-zero lattice spacing. Listed for \PQCPT\ and \CPT\ are the number of linearly independent
operators contributing to the masses of the pion ($m_\pi$), nucleon ($M_N$), and delta ($M_\D$). These operators
are classified as $\cO(a)$, $\cO(m_Q \,a)$, or $\cO(a^2)$. For each theory
the quarks are grouped by species and for partially quenched theories are grouped by 
pairs of valence and sea quark species. The species are labeled W for Wilson, and GW for Ginsparg-Wilson.}
%\begin{ruledtabular}
\begin{tabular}{c | c c c | c c c | c c c }
\PQCPT\ & \multicolumn{3}{c|}{$m_\pi$} & \multicolumn{3}{c|}{$M_N$} & \multicolumn{3}{c}{$M_\D$} \\
$(v)$ $(s)$ & $\; \cO(a)$ & $\; \cO(m_Q \, a) \;$ & $\; \cO(a^2) \;$ 
	    & $\; \cO(a)$ & $\; \cO(m_Q \, a) \;$ & $\; \cO(a^2) \;$  
            & $\; \cO(a)$ & $\; \cO(m_Q \, a) \;$ & $\; \cO(a^2) \;$  \\
\hline
W W   & $1$ & $2$ & $1$ & $1$ & $2$ & $2$ & $1$ & $2$ & $3$ \\
W GW  & $1$ & $1$ & $1$ & $1$ & $2$ & $2$ & $1$ & $2$ & $3$ \\
GW W  & $0$ & $1$ & $0$ & $1$ & $2$ & $2$ & $1$ & $2$ & $3$ \\
GW GW & $0$ & $0$ & $0$ & $0$ & $0$ & $2$ & $0$ & $0$ & $3$ \\
\multicolumn{10}{c}{} \\
\hline
$SU(2)$ \CPT\ & $\; \cO(a)$ & $\; \cO(m_Q \, a) \;$ & $\; \cO(a^2) \;$  
	      & $\; \cO(a)$ & $\; \cO(m_Q \, a) \;$ & $\; \cO(a^2) \;$  
              & $\; \cO(a)$ & $\; \cO(m_Q \, a) \;$ & $\; \cO(a^2) \;$  \\
W      & $1$ & $1$ & $1$ & $1$ & $1$ & $2$ & $1$ & $1$ & $3$ \\
GW     & $0$ & $0$ & $0$ & $0$ & $0$ & $2$ & $0$ & $0$ & $3$ \\
\end{tabular}
%\end{ruledtabular}
\label{t:summary}
\end{table}

\section{\label{s:summy}Summary}

Above we have extended heavy baryon chiral perturbation theory for the Wilson action to $\cO(a^2)$. 
In our power counting, we have included all operators that are at least of $\cO(\e^4)$, 
this includes operators of $\cO(m_Q \, a)$ and $\cO(a^2)$. We have considered partially quenched $SU(4|2)$ 
with a mixed action, as well as $SU(2)$ in Appendix~\ref{s:2}.
To this order in the combined lattice spacing and
chiral expansion, we saw the necessity for the introduction of a large number of new operators. For the case
of a partially quenched theory with a mixed lattice action, there are well over one hundred operators in the
free Lagrangian containing the spin-$\frac{1}{2}$ and spin-$\frac{3}{2}$ baryons.

Despite the introduction of a large number of terms with undetermined LECs, observables calculated to $\cO(a^2)$ 
depend on still relatively few independent parameters. This is because many of the operators introduced
for mixed lattice actions act differently depending on the number of sea quarks contained in particular states of the multiplet. 
Each level of the multiplet must be treated differently in the effective theory because there are no symmetry transformations
between the valence and sea sectors of the theory. This distinction is important if the level-dependent 
operators act on baryons within loops and such a situation only occurs at higher orders in the chiral expansion. 
Moreover, most of the remaining operators for baryons which consist of only valence quarks  
are not independent at leading order and hence 
cannot have their LECs disentangled from lattice data. To make these points clear, we have calculated 
the finite lattice-spacing corrections to baryon masses in various theories. In Table~\ref{t:summary},
we summarize the number of free parameters due to lattice spacing artifacts 
that enter into expressions for baryon masses and can be determined from lattice data.  
The table is separated for the pion, nucleon, and delta into contributions from independent $\cO(a)$, 
$\cO(m_Q \, a)$ and $\cO(a^2)$ operators.

There are some further points to observe about the baryon sector at $\cO(a^2)$. 
While there are considerably more operators in the partially quenched theories for mixed actions
as compared to the unquenched theories, the number of independent parameters entering 
into the determination of baryon masses from lattice data is the same for $\cO(a)$ and $\cO(a^2)$ operators;
there is only one additional $\cO(m_Q \, a)$ parameter which stems from the different sea quark mass. 
Nonetheless, the computational benefits of partially quenching are not hindered by the explosion 
in the number of new baryon operators because most are not independent to $\cO(\e^4)$.

On the other hand, based on studies in the meson sector~\cite{Bar:2003mh}, one would expect
the number of free parameters to be reduced when considering theories with Ginsparg-Wilson
valence quarks and Wilson sea quarks compared to the Wilson action, \emph{cf}. Table~\ref{t:summary}. 
While there is a clear and sizable reduction in the number 
of baryon operators for actions employing Ginsparg-Wilson valence quarks and Wilson sea quarks
compared to the unmixed Wilson action, 
there is no reduction in the number of free parameters involved in the baryon masses. 
The only reduction in the number of free parameters comes about in simulations employing Ginsparg-Wilson 
quarks in both the valence and sea sectors. Another notable difference for heavy baryon fields is that operators which
reduce the $O(4)$ symmetry to the hypercubic group are present in the effective theory. Such operators were suppressed 
in the pseudoscalar meson sector because the absence of a large mass scale implies
the pseudoscalar momenta are small. For heavy baryon fields, however, there is a large mass scale and the momenta
are automatically large, thus such operators are not suppressed when multiplied by powers of the lattice spacing.  
Similar operators must then enter at $\cO(\e^4)$ for other heavy particles. 
We have spelled out the form of $O(4)$ breaking corrections for particles of spin less than two in Appendix~\ref{s:RPI}.

In this work we have seen that near the continuum limit, the spin-$\frac{1}{2}$ baryon self energy has the 
behavior\footnote{For GW quarks, the coefficients $\a$, $\b$, and $\gamma$ are identically zero.}
\begin{equation}
M(a) = M(0) + \a \, a + \b \, a \, m_q \, \log m_q + \g \, a \, m_q + \d \, a^2 
+ \epsilon \, \ol u(p) \, v_\mu v_\mu v_\mu v_\mu \, u(p) + \ldots
,\end{equation}
where the omitted terms are of order $\e^5$ and higher. 
In writing the above expression, we have assumed for simplicity that one fine tunes the quark mass so that the pion mass vanishes
in the chiral limit. In this case there are only polynomial corrections in $a$. The influence of the $O(4)$ breaking term  
on the baryon energy was determined in Sec.~\ref{s:nm}.
For spin-$\frac{3}{2}$ baryons, the behavior of the self energy has the same form, however there is an additional 
contribution to the self energy of the form
\begin{equation}
\ol u_\mu(p) \, \ol \Sigma_{\mu \mu} \, u_\mu(p) = \omega \, a^2 \, \ol u_\mu(p) \, v_\mu v_\mu \, u_\mu(p) + \ldots
,\end{equation}
and we have detailed the form of this correction in Sec.~\ref{s:dm}. 
A similar such additional term is present for vector mesons as well.

Knowledge of the quark mass dependence of baryon observables is crucial to perform the
chiral extrapolation of lattice data, extract physical LECs, and make predictions for QCD. 
Artifacts of approximating spacetime on a discrete lattice make the chiral extrapolation 
more challenging because of the introduction of additional error. With \CPT\ and \PQCPT\ 
formulated for the Symanzik action, one can parametrize the dependence on the lattice
spacing and considerably reduce the uncertainty surrounding lattice artifacts. 
The formalism set up here is straightforward to extend to other Wilson-type fermion actions,
twisted mass QCD~\cite{Walker-Loud:2005bt}, for example. 
On the other hand, due to the proliferation of operators, it is doubtful that partially 
quenched baryon staggered \CPT\ would provide much analytic insight at $\cO(a^2)$.

\begin{acknowledgments}
We are indebted to Martin Savage for alerting us to \cite{Savage}. 
We would also like to acknowledge Shailesh Chandrasekharan, Will Detmold,  and Andr\'e Walker-Loud for helpful discussions.
This work is supported in part by the U.S.\ Department of Energy under 
Grant No.\ DE-FG02-96ER40945, and we thank the Institute for Nuclear Theory
at the University of Washington for its hospitality during the final stages of this work. 
\end{acknowledgments}

\appendix

\section{Baryon masses to $\cO(a^2)$ in $SU(2)$ \CPT} \label{s:2}

In this Appendix, we detail the case of finite lattice-spacing corrections 
to baryon masses in $SU(2)$ \CPT. First we write down all finite lattice spacing terms that arise
to $\cO(\e^4)$ from the Symanzik Lagrangian. Next we determine the corrections
from these operators to the masses of the nucleon and delta.

The form of the Symanzik Lagrangian for $SU(2)$ flavor is mainly the same 
as in the main text. Of course there is no separation of the 
theory into valence and sea sectors.  The mass matrix for two light flavors
in the isospin limit is given by $m_Q = \diag (m_u , m_u)$, while the Wilson matrix appears as
$w_Q = \diag (w_v, w_v)$. The subscript $v$ now has no particular significance 
in $SU(2)$ \CPT\ other than to maintain the definition of $w_v$ used in above.

At zero lattice spacing and zero quark mass, two flavor QCD has an $SU(2)_L \otimes SU(2)_R \otimes U(1)_V$
chiral symmetry that is broken down to $SU(2)_V \otimes U(1)_V$. \CPT\ is the low-energy 
effective theory that emerges from perturbing about the physical vacuum. 
The pseudo-Goldstone bosons can be described by a Lagrangian that takes into account
the two sources of chiral symmetry breaking and is given in Eq.~\eqref{eqn:Lchi}, 
with the exception that $\Phi$ is an $SU(2)$ matrix containing just the familiar pions, and 
the $\str$'s are now $\tr$'s. Their masses are given in Eq.~\eqref{eqn:mqq}.

In $SU(2)$ \CPT, the delta baryons are contained in the flavor tensor $T^{ijk}_\mu$ which 
is embedded in the \PQCPT\ tensor $\cT^{ijk}_\mu$ simply as $\cT^{ijk}_\mu = T^{ijk}_\mu$,  when all indices
are restricted to $1$--$2$. Consequently the form of the \CPT\ Lagrangian for the delta fields 
has a form very similar to that of \PQCPT.
The nucleons, however, are conveniently described by an $SU(2)$ doublet
\begin{equation}
N = 
\begin{pmatrix}
p \\
n
\end{pmatrix}
.\end{equation}
These states are contained in the \PQCPT\ flavor tensor $\cB^{ijk}$
as~\cite{Beane:2002vq}
\begin{equation}
\cB^{ijk} = \frac{1}{\sqrt{6}} 
\left( 
\e^{ij} N^{k} + \e^{ik} N^{j}
\right)
,\end{equation}
when all of the indices are restricted to $1$--$2$. Consequently operators involving the nucleons
will appear differently in $SU(2)$.

To $\cO(\e^2)$ the free Lagrangian for the nucleons and deltas reads
\begin{eqnarray}
\cL 
&=& 
i \ol N  v \cdot D N 
- 
2 \sigma
\ol N N \tr (\cM_+)
- 
2 \sigma_w 
\ol N N \tr ( \cW_+ ) 
\notag \\
&& +
i \ol T_\mu v \cdot D T_\mu 
+ 
\D \ol T_\mu T_\mu 
- 
2 \ol \sigma \, \ol T_\mu T_\mu \tr (\cM_+)
-
2 \ol \sigma_w \ol T_\mu T_\mu \tr (\cW_+)
.\end{eqnarray}
The LECs for the nucleons and deltas in $SU(4|2)$ are related by the matching equations
\begin{eqnarray}
\sigma     &=&   \frac{1}{2} ( \a_M + \b_M) + \sigma_M, \notag \\
\ol \sigma &=& - \frac{1}{2} \gamma_M + \ol \sigma_M 
,\end{eqnarray}
and completely analogous equations that relate 
$\sigma_w$, and $\ol \sigma_w$ to the \PQCPT\ parameters 
$\a_W$, $\b_W$, $\sigma_W$, $\gamma_W$, and $\ol \sigma_W$. 
The interaction Lagrangian of $SU(2)$ appears as
\begin{equation}
\cL = 2 g_A \ol N S \cdot A N + 2 g_{\D N} \left( \ol T_\mu A_\mu N + \ol N A_\mu T_\mu \right)
- 2 g_{\D \D} \ol T_\mu S \cdot A T_\mu  
.\end{equation}
The relation of the parameters appearing in the \PQCPT\ Lagrangian in the main text to those in \CPT\
can be found from matching. One finds \cite{Beane:2002vq}: $g_A = \frac{2}{3} \a_M - \frac{1}{3} \b_M$, 
$g_{\D N} = - \cC$, and $g_{\D \D} = \cH$.

The first lattice spacing corrections at $\cO(\e^4)$ arise from the $\cO( m_Q \, a)$ operators. In degenerate $SU(2)$, we have
only two such baryon operators
\begin{eqnarray}
\cL 
&=&
- \frac{1}{ \L_\chi} \Bigg[
n^{WM} \, \ol N N \, \tr( \cW_+) \, \tr (\cM_+)  
-
t^{WM} \, \ol T_\mu T_\mu \, \tr( \cW_+) \, \tr (\cM_+)
\Bigg]
.\end{eqnarray}
The remaining terms are of order $\cO(a^2)$ and appear in the Lagrangian
\begin{eqnarray}
\cL &=&
- \frac{1}{\L_{QCD}} \Bigg[
n^W \, \ol N N \, \tr(\cW_+) \tr (\cW_+ ) 
-
t^W \, \ol T_\mu T_\mu \, \tr(\cW_+) \tr (\cW_+ ) 
\Bigg] \notag \\
&& - 
a^2 \L_{QCD}^3 \Big[ n \, \ol N N + n^v \ol N v_\mu v_\mu v_\mu v_\mu N 
- t \, \ol T_\mu T_\mu - t^v \, \ol T_\nu v_\mu v_\mu v_\mu v_\mu T_\nu 
- t^{\ol v} \, \ol T_\mu v_\mu v_\mu T_\mu \Big]
.\end{eqnarray}
The coefficients of operators in the $SU(2)$ theory are contained in $SU(4|2)$, but 
because $SU(4|2)$ contains considerably more operators, the matching yields 
algebraically cumbersome expressions that relate the LECs.

Having written down all of the relevant operators for the nucleon and delta masses, we now calculate their lattice
spacing corrections. We do not write down the loop contributions from $a$-independent operators because the only 
modification necessary is to include the lattice spacing dependence of the meson masses that appears in Eq.~\eqref{eqn:mqq}. 
For the mass of the nucleon, at $\cO(\e^2)$ we find
\begin{equation}
M_N^{(2)} = 4 \sigma \, m_u  
+ 4 a \L_{QCD}^2 \sigma_w \, w_v
,\end{equation}
The finite lattice spacing corrections to $M_N^{(4)}$ read
\begin{eqnarray}
\delta M_N^{(4)} 
&=& 
4 w_v \frac{a \L_{QCD}^2}{\L_\chi^2} \Big\{
\L_\chi \, n^{WM} m_u
- 
3 \sigma_w \cL(m_\pi, \mu)
+
4 g_{\D N}^2 (\ol \sigma_w - \sigma_w ) 
\left[ \cJ(m_\pi, \D, \mu) + m_\pi^2  \right]
\Big\}
\notag \\
&& +
a^2 \L_{QCD}^3 
\Big( n + 4 n^W w_v + n^v \ol u(p) v_\mu v_\mu v_\mu v_\mu \, u(p) \Big)
.\end{eqnarray}
For the deltas, similar calculations yield the $\cO(\e^2)$ result
\begin{equation}
M_\D^{(2)} 
= 
- 
4 \ol \sigma \, m_u
-
4 a \L_{QCD}^2 
\ol \sigma_w \, w_v 
\label{eq:Tmass6}
.\end{equation}
At $\cO(\e^4)$ the remaining finite lattice spacing corrections to delta mass are
\begin{eqnarray}
\delta M_\D^{(4)} 
&=& 
4 w_v \frac{a \L_{QCD}^2}{\L_\chi^2} \Big[ 
\L_\chi  \, t^{WM} m_u
+ 
3 \ol \sigma_w \cL(m_\pi, \mu)
+ 
g_{\D N}^2 (\ol \sigma_w -  \sigma_w) \cJ(m_\pi , - \D, \mu)
\Big]
\notag \\
&& +
a^2 \L_{QCD}^3 
\Big( 
t + 4 t^{W} \, w_v + t^v \ol u_\nu(p) v_\mu v_\mu v_\mu v_\mu \, u_\nu(p) + t^{\ol v} \ol u_\mu(p) v_\mu v_\mu \, u_\mu(p)
\Big)
.\label{eq:Tmass7}
\end{eqnarray}

\section{$O(4)$ breaking operators} \label{s:RPI}

Above we addressed $\cO(a^2)$ corrections in the baryon sector. At this order
we saw that $O(4)$ breaking operators are present in the chiral effective theory. 
In this Appendix, we detail the form of these corrections in heavy particle effective
theories for particles of spin less than two.

We consider the simplest case first, that of a heavy scalar field $\phi$. 
At $\cO(a^2)$, we have an $O(4)$ breaking operator of the form
\begin{equation} \label{eq:scalar}
\cL = a^2 c_0 \, \phi^\dagger \, v_\mu v_\mu v_\mu v_\mu \, \phi
.\end{equation}
We can either treat the four-velocity $v^\mu$ as the physical velocity of the $\phi$ field, or 
consider the four-velocity to be fixed in the rest frame, for example. Clearly when
the particle is at rest, both descriptions are identical; and with $v_\mu = (0,0,0,i)$, 
we obtain a trivial $a^2$ dependent shift to the $\phi$ mass. Away from rest the effects 
of $O(4)$ breaking are no longer trivial. 

First let us consider $v_\mu$ to be the physical velocity of the $\phi$. In terms of 
the particle's spatial momentum $\bm{p}$, we have $v_\mu = (\gamma \bm{p}/M, i \gamma)$, 
where $M$ is the mass of the $\phi$ and $\gamma$ is the Lorentz-Fitzgerald contraction factor. 
In a non-relativistic expansion, we can express the particle's four-velocity in terms of the 
spatial momentum. This enables us to write
\begin{equation} \label{eq:vvvv}
\sum_{\mu = 1}^{4} v_\mu v_\mu v_\mu v_\mu  
= 
1 
+ 
2 \frac{\bm{p}^2}{M^2} 
+ 
\frac{1}{M^4} \left( \sum_{i=1}^{3} p_i p_i p_i p_i + \bm{p}^4 \right)
+ 
\cO(M^{-6})
.\end{equation}
Thus the operator in Eq.~\eqref{eq:scalar}, besides shifting the mass by $a^2 c_0$, modifies
the particle's kinetic energy as well as the relativistic correction to the kinetic energy. 
The energy of the $\phi$ is now a cubic invariant function of the spatial momentum.
The breaking of $O(3)$ rotational symmetry for a particle's energy-momentum relation
is, however, small for non-relativistic momenta.

Extending the above analysis to non-zero spin involves boosting the various spin
wavefunctions. On the other hand, if one treats the four-velocity as fixed and one can 
derive a string of fixed coefficient operators using reparameterization invariance ~\cite{Luke:1992cs}.
For a fixed four-velocity, we write
the heavy scalar momentum $P$ as $P_\mu = M v_\mu + p_\mu$. This decomposition is arbitrary as one can make 
a reparameterization of the form
\begin{equation}
\begin{cases}
v_\mu \to v_\mu +  q_\mu / M \notag \\
k_\mu \to k_\mu - q_\mu
\end{cases}
\label{eq:RPS}
.\end{equation}
Accordingly the field $\phi(x)$ changes to $e^{i q_\mu x_\mu} \phi(x)$, and
to maintain the normalization of the new four-velocity, we require $v \cdot q = - q^2 / (2 M)$.
The physics is unchanged by this transformation and this is referred to as reparameterization invariance. 
In order for the Lagrangian in Eq.~\eqref{eq:scalar} to be reparameterization invariant (RPI), there must 
be additional operators with fixed coefficients. These can all be written compactly in the manifestly
RPI form
\begin{equation} \label{eq:scalarRPI}
\cL = a^2 c_0 \, \phi^\dagger \, 
\left(v_\mu + \frac{i D_\mu}{M}\right)
\left(v_\mu + \frac{i D_\mu}{M}\right)
\left(v_\mu + \frac{i D_\mu}{M}\right)
\left(v_\mu + \frac{i D_\mu}{M}\right)
\phi
.\end{equation}
%where $\partial_{\perp,\mu} = \partial_\mu + v_\mu v \cdot \partial$. 
Expanding out the terms of Eq.~\eqref{eq:scalarRPI} and acting between states of residual momentum $p_\mu$
taken in the non-relativistic limit, we recover the result of Eq.~\eqref{eq:vvvv}.

For a heavy spin-$1/2$ particle $\psi$, we have the $O(4)$ breaking Lagrangian
\begin{equation} \label{eq:half}
\cL = a^2 c_{1/2} \ol \psi \, v_\mu v_\mu v_\mu v_\mu \, \psi
,\end{equation}
which is not RPI. Unlike the scalar case, we have a further constraint on the field
due to the spin degrees of freedom. To maintain the condition $- i \rlap \slash v \psi = \psi$, 
the field $\psi(x)$ must change 
%to $e^{i q_\mu x_\mu} [ 1 - i \rlap \slash q / (2 M)] \psi(x)$ 
under the reparameterization in Eq.~\eqref{eq:RPS}.  
To order $M^{-2}$, we have reparameterization invariant spinor $\Psi$ given by~\cite{Manohar:1997qy} 
\begin{equation}
\Psi = \left( 1 - \frac{\Dslash_\perp}{2 M } - \frac{D^2}{8 m^2} \right) \psi
,\end{equation}
where $D_{\mu,\perp} = D_\mu + v_\mu v_\nu D_\nu$.  
Since the bilinear $\ol \Psi \Psi = \ol \psi \psi + \cO(M^{-3})$, the RPI form of Eq.~\eqref{eq:half} is
\begin{equation} \label{eq:halfRPI}
\cL = a^2 c_{1/2} \ol \psi 
\left(v_\mu + \frac{i D_\mu}{M}\right)
\left(v_\mu + \frac{i D_\mu}{M}\right)
\left(v_\mu + \frac{i D_\mu}{M}\right)
\left(v_\mu + \frac{i D_\mu}{M}\right)
\psi
,\end{equation}
and hence the correction to the energy to $\cO(M^{-2})$ has the same form as in Eq.~\eqref{eq:vvvv}.
If we were to write the covariant spinor to fourth order, we could ascertain the coefficient of
the $\bm{p}^4/M^4$ term in Eq.~\eqref{eq:vvvv} for a spin-$1/2$ particle. 
The $O(3)$ breaking term, however, is unchanged.

The correction for a heavy vector particle, the rho meson $\rho_\mu$ for example, is deduced similarly.
The $\cO(a^2)$ Lagrangian contains two terms 
\begin{equation} \label{eq:rho}
\cL = a^2 c_1 \, \rho_\nu^\dagger \, v_\mu v_\mu v_\mu v_\mu \, \rho_\nu 
+ 
a^2  \ol c_1 \, \rho^\dagger_\mu \, v_\mu v_\mu \, \rho_\mu 
,\end{equation}
and hence there is a new type of contribution to the energy proportional to $\epsilon^*_\mu (p) \, v_\mu v_\mu \, \epsilon_\mu (p)$, 
where $\epsilon_\mu(p)$ is the rho's polarization vector. This correction vanishes for a rho at rest. 
The terms in Eq.~\eqref{eq:rho} are not RPI. To deduce the RPI form of the Lagrangian, we must now also 
maintain the constraint $v_\mu \rho_\mu = 0$. We can build the RPI Lagrangian from RPI vector field $R_\mu$.
%
%transform the field $\rho_\mu(x)$ according to $e^{- i q_\mu x_\mu} (\d_{\mu \nu} - v_\mu q_\nu / M ) \rho_\nu(x)$
%in order to preserve $v_\mu \rho_\mu = 0$ under the reparametrization Eq.~\eqref{eq:RPS}.
To $\cO(M^{-2})$, we have 
\begin{equation} \label{eq:rhocov}
R_\mu = \left(\d_{\mu \nu} - v_\mu \frac{i D_\nu }{M} + \frac{D_\mu D_\nu}{M^2}\right) \rho_\nu
,\end{equation}
and hence the RPI form of the Lagrangian in Eq.~\eqref{eq:rho} can be deduced from combinations of $R^\dagger_\mu \Gamma R_\mu$. 
Because we are addressing the corrections at tree level, the states are on-shell and 
the additional terms in Eq.~\eqref{eq:rhocov} vanish. In loops the off-shell degrees of freedom
propagate and the additional terms are necessary. 
The on-shell RPI Lagrangian is thus given by
\begin{eqnarray} \label{eq:one}
\cL &=& a^2 c_1 \, \rho_\nu^\dagger 
\left(v_\mu + \frac{i D_\mu}{M}\right)
\left(v_\mu + \frac{i D_\mu}{M}\right)
\left(v_\mu + \frac{i D_\mu}{M}\right)
\left(v_\mu + \frac{i D_\mu}{M}\right)
\rho_\nu \notag \\
&& + 
a^2 \ol c_1 \, \rho_\mu^\dagger 
\left(v_\mu + \frac{i D_\mu}{M}\right)
\left(v_\mu + \frac{i D_\mu}{M}\right)
\rho_\mu
.\end{eqnarray}

We can use the above Lagrangian to determine the $\cO(a^2)$ corrections
to the vector particle's dispersion relation. Let us denote $M(a)$ 
as the particle mass including lattice discretization contributions. 
One such contribution is $a^2 c_1$ that arises from the above Lagrangian. There 
are many others, however, that we have not spelled out. Modification of the
dispersion relation only arises from the terms in Eq.~\eqref{eq:one}.
Utilizing the non-relativistic expansion for the momentum $\bm{p}$ 
and explicit forms of the polarization vectors, 
we arrive at the energy of a spin-one particle
\begin{equation}
E^\l_{\bm{p}} = M(a) + \frac{\bm{p}^2}{2 M(a)} \left( 1 + \frac{4 a^2 c_1}{M(0)}\right) +
\frac{a^2 \ol c_1}{2 M(0)^2} \left[ |\l| \bm{p}_\perp^2 + 2 (|\l| - 1) \bm{p}_\l^2 \right]
,\end{equation}
where $M(0)$ is the rho mass in the continuum limit, $\lambda = \pm 1, 0$ is the 
spin along an axis, $\bm{p}_\l$ is the momentum along that axis, 
and $\bm{p}_\perp$ is the momentum transverse to that axis.
For a vector particle of given polarization, the energy-momentum relation breaks
$O(3)$ rotational invariance at $\cO(a^2 M^{-2})$. Notice for an unpolarized state,
the rotational symmetry breaking from this term is averaged out and then does not enter until $\cO(a^2 M^{-4})$
as with scalar and spin-$1/2$ particles.

Finally we address the corrections for the case of a spin-$3/2$ particle represented by a Rarita-Schwinger
field $\psi_\mu$. At $\cO(a^2)$ terms that break $O(4)$ appear in the Lagrangian as
\begin{equation} \label{eq:3/2}
\cL = 
a^2 c_{3/2} \, \ol \psi_\nu \, v_\mu v_\mu v_\mu v_\mu \, \psi_\nu
+
a^2 \ol c_{3/2} \, \ol \psi_\mu \, v_\mu v_\mu \, \psi_\mu 
.\end{equation}
To deduce the corrections to the energy, we cast this Lagrangian into its RPI form. 
Under a reparameterization, one must maintain the conditions $- i \rlap \slash v \psi_\mu = \psi_\mu$ and $v_\mu \psi_\mu = 0$. 
The RPI field $\Psi_\mu$ is a combination of terms from RPI invariant spinor and vector fields and is given by
\begin{equation}
\Psi_\mu 
= 
\left[
\d_{\mu \nu} - \frac{1}{2 M} 
\left( 
\Dslash {}_\perp + i v_\mu D_\nu
\right) 
- \frac{1}{8 M^2}
\left( 
D^2 - 4 i v_\mu \Dslash {}_\perp D_\nu
- 8 D_\mu D_\nu 
\right)
\right] 
\psi_\nu
,\end{equation}
up to $\cO(M^{-2})$. Half of the terms in the RPI spin-$3/2$ field vanish on shell. The remaining terms 
are identical to those for a spin-$1/2$ particle. Thus for an on-shell particle, we have
$\ol \Psi_\mu \Psi_\mu = \ol \psi_\mu \psi_\mu + \cO(M^{-3})$. Thus the on-shell RPI Lagrangian
corresponding to Eq.~\eqref{eq:3/2} is given by
\begin{eqnarray}
\cL &=& 
a^2 c_{3/2} \, \ol \psi_\nu 
\left(v_\mu + \frac{i D_\mu}{M}\right)
\left(v_\mu + \frac{i D_\mu}{M}\right)
\left(v_\mu + \frac{i D_\mu}{M}\right)
\left(v_\mu + \frac{i D_\mu}{M}\right)
\psi_\nu \notag \\
&& + 
a^2 \ol c_{3/2} \, \ol \psi_\mu 
\left(v_\mu + \frac{i D_\mu}{M}\right)
\left(v_\mu + \frac{i D_\mu}{M}\right)
\psi_\mu \label{eq:RS}
.\end{eqnarray}
As with the spin-one case, we can evaluate the corrections to the spin-$3/2$ energy in the non-relativistic limit
using the Lagrangian Eq.~\eqref{eq:RS} and explicit forms of the Rarita-Schwinger vectors. The result for the $\D$ mass
is presented in the main text, see Eq.~\eqref{eq:new}.

\bibliography{hb}

\end{document}